%% file: PenaltyPaper.tex
\documentclass[12pt]{article}
\usepackage[left = 2.54cm, right = 2.54cm, top = 2.54cm, bottom = 2.54cm]{geometry} 
\usepackage{setspace}
\usepackage{graphicx}
\usepackage{amsfonts} 
\usepackage{amsmath}
\usepackage{float}
\restylefloat{table}
\usepackage{natbib} 
\usepackage{placeins}
\usepackage{caption}
\usepackage{subcaption} 
\usepackage{amsthm}
\numberwithin{equation}{section}
\usepackage[ruled]{algorithm2e}
\usepackage{titlepic}
\DeclareMathAlphabet{\mathpzc}{OT1}{pzc}{m}{it}

\usepackage{color}
\usepackage[toc,page]{appendix}
\usepackage{fancyhdr} 
\usepackage{url}
\usepackage{xcolor}
\usepackage{amsmath}
\usepackage{authblk}

\newtheorem{theorem}{Theorem}[section]

 \title{Efficient penalty search for multiple changepoint problems}
\author[1$\dag$]{Kaylea Haynes}
\author[2]{Idris A. Eckley}
\author[2]{Paul Fearnhead} 
\affil[1]{STOR-i Centre for Doctoral Training, Lancaster University}
\affil[2]{Department of Mathematics and Statistics, Lancaster University}
\affil[$\dag$]{Correspondence: k.haynes1@lancaster.ac.uk}

\begin{document}
\singlespacing
\maketitle 
  
\begin{center}
 {\bf Abstract}
\end{center}


In the multiple changepoint setting, various search methods have been proposed which involve optimising either a constrained or penalised cost function over possible numbers and locations of changepoints 
using dynamic programming.  Such methods are typically computationally intensive. Recent work in the penalised optimisation setting has focussed on developing a pruning-based approach 
which gives an improved computational cost that, under certain conditions, is linear in the number of data points.  
Such an approach naturally requires the specification of a penalty to avoid under/over-fitting.  Work has been undertaken to identify the appropriate penalty choice for data generating processes with 
known distributional form, but in many applications the model assumed for the data is not correct and these penalty choices are not always appropriate. 
Consequently it is desirable to have an approach that enables us to compare segmentations for different
choices of penalty.  To this end we present a method to obtain optimal changepoint segmentations of data sequences for all
penalty values across a continuous range.  This permits an evaluation of the various segmentations to identify a suitably parsimonious penalty choice. The computational complexity of this approach can be 
linear in the number of data points and linear in the difference between the number of changepoints in the optimal segmentations for the smallest and largest penalty values. This can be orders of magnitude faster
than alternative approaches that find optimal segmentations for a range of the number of changepoints.





\input{intro_IE}

\input{background3}
\input{method2}
\input{simulation3}
\input{discussion}

{\bf Acknowledgements:} This research was supported by EPSRC grant EP/K014463/1. Haynes gratefully acknowledges the financial support of DSTL and also the EPSRC Centre for Doctoral Training in Statistics and Operational Research in partnership with Industry.  

\bibliography{MyRefs,Refs2}

\end{document}

%% file: intro_IE.tex


\section{Introduction}
High resolution data sensors are common-place in the devices which we use in our day to day lives. Consequently we are now able to record and store more data than ever before.  This has resulted in a resurgence of interest in a number of different inference areas, not least of which is changepoint analysis. See for example contributions in finance \citep{Aggarwal1999}, computer science \citep{Yan2008}, and environmental disciplines including oceanography \citep{Killick2013} and climatology \citep{Reeves2007}. 

Changepoints are considered to be those points in a data sequence where we observe a change in the statistical properties, such as a change in mean, variance or distribution. Assume we have data, $y_1,\ldots,y_n$, that has been ordered based on some covariate information, for example by time or by position along a chromosone. For clarity we will assume we have time-series data in the following.  Our time-series will have $m$ changepoints with locations $\tau_{1:m} = (\tau_1,...,\tau_m)$ where each $\tau_i$ is an integer between 1 and $n-1$ inclusive. We assume that $\tau_i$ is the time of
the $i$th changepoint, so that $\tau_1 < \tau_2 < ... < \tau_m$. We set $\tau_0 = 0$ and $\tau_{m+1} = n$ so that  the changepoints split the data into $m+1$ segments 
with the $i$th segment containing the data points $y_{(\tau_{i-1} + 1):\tau_{i}}=(y_{\tau_{i-1}+1},\ldots,y_{\tau_i})$. 

One common approach to changepoint detection is to define a cost for a given segmentation of the data. 
Typically this cost is based on first defining a segment-specific cost function, which we denote as $\mathpzc{C}(y_{s+1:t})$ for a segment which contains data-points $y_{s+1:t}$. We then sum this segment-specific cost function over the $m+1$ segments.  A natural way to then estimate the number and position of the changepoints would be to minimise the resulting cost over all segmentations. 

However, for many cost functions, this results in overfitting since adding a changepoint always reduces the overall cost. 
There are two potential approaches to avoiding such overfitting.  The first of these would be to use knowledge of the application to constrain the optimisation by fixing the maximum number of changepoints that can be found.  
The corresponding {\em constrained minimisation problem} is:
\begin{align}
Q_m(y_{1:n}) = \min_{\tau_{1:m}} \left \{ \sum_{i=1}^{m+1} [\mathpzc{C}(y_{(\tau_{i-1} + 1) : \tau_i})] \right \},
\label{eqn:const} 
\end{align}  
with the best segmentation with $m$ changepoints being the one that attains the minimum.  If the number of changepoints is unknown then the number of changes, $m$, is often estimated by solving
\begin{align} \label{eq:pen_cp}
\min_m  \left \{  Q_m(y_{1:n}) + f(m) \right \},
\end{align}  
where $f(m)$ is a suitably chosen penalty term that increases with $m$. \\ 

If $f(m)$ is a linear function, that is $f(m) = (m+1)\beta$ with $\beta>0$, then we can jointly estimate the number and the position of the changepoints by solving a {\em penalised minimisation problem}:
\begin{align}
Q(y_{1:n}, \beta) = \min_{m,\tau_{1:m}} \left \{  \sum_{i=1}^{m+1} [\mathpzc{C}(y_{(\tau_{i-1} + 1) : \tau_i}) + \beta]\right \}, \label{eqn:pen_like}
\end{align}
again with the estimated segmentation being the one that attains the minima. This second approach, of directly minimising (\ref{eqn:pen_like}) is computationally faster than solving the constrained penalisation
problem for a range of the number of changepoints, and then minimising (\ref{eq:pen_cp}). However it requires a choice of penalty constant, $\beta$.


The choice of penalty constant can have an important impact on the accuracy of the segmentation estimate that we obtain. The process of appropriately selecting the penalty value is not usually straightforward. 
Many authors have looked at different choices of penalties. If we let $p$ denote the number of additional parameters introduced by adding a changepoint, then popular examples used frequently in the literature include 
$\beta = 2p$ \cite[Akaike's Information Criterion;][]{Akaike1974};   $\beta = p \log n$
\cite[Schwarz's Information Criterion;][]{Schwarz1978}; and $\beta = 2p \log\log n$ \citep{Hannan1979}. 
In all cases, some assumptions are made about the underlying data generating process which gives rise to the data. Unfortunately, in practice one may not know \emph{a priori} whether the data can be assumed to arise from such a process. In such cases, there is a potential for the modelling assumptions associated with a particular criterion to be violated. As we shall see later, one therefore risks over/under-fitting the data.

Our main contribution in this article is to propose a method for using a range of values of $\beta$ in changepoint detection instead of selecting a single value.  This approach allows us to compare the resulting segmentations for different penalty values. Our method uses the relationship between minimising a penalised cost function in relation to a cost function which is constrained by the number of segments.  That is, we can find the corresponding constrained cost with $m^*$ changepoints which results in the same segmentation that we get if we have a penalised cost with penalty value $\beta^*$.  Using this link we find that solving the penalised optimisation problem with some values of the penalty result in the same number of changepoints and thus we propose a method which only solves the optimisation problem with penalty values which result in different solutions and use these to provide the number of changes for all possible penalty values.   We show how the computational cost of the method can be 
improved by reusing common results between different penalty values. We show in Section \ref{sec:sims} that this method finds optimal segmentations of the data with different number of changes faster than Segment Neighbourhood Search. 

The article is organised as follows.  In Section \ref{sec:back} we introduce the changepoint model and review various ways of detecting multiplechanges using both a constrained and a penalised approach.  In Section \ref{sec:method} we propose our method for running the detection algorithms over a range of penalty values.  This method utilises a link between the constrained and the penalised approaches.  Within this section we discuss how we can make improvements to the computational cost by recycling common results within two popular search methods.  Our method will be demonstrated in simulation studies and real data examples in Section \ref{sec:sims} and Section \ref{real_data}.

%% file: background3.tex
%
%
%
%

\section{Background}\label{sec:back}

\subsection{Segment Costs}
To define the cost of a segmentation used in either the constrained or penalised optimisation problems introduced above, we need to specify a segment-specific cost. A common approach, used for example in penalised
likelihood \cite[]{Braun/Muller:1998} and minimum description length \cite[]{Davis2006} methods, is to introduce a model for the data within a segment. This will define a log-likelihood for the data that depends on a 
segment-specific parameter.  The cost can then be chosen proportional to minus the maximum of this log-likelihood, where we maximise out the segment-specific parameter. The form of this cost
will then depend on both modelling assumptions about the distribution of the data points, and also the type of change that we are attempting to detect. Whilst there are other approaches to defining costs, in many cases
these are equivalent to a cost based on an appropriate likelihood model, for example note the link between a change in mean for a Gaussian model and a square error cost below.


To make this idea concrete, consider the following setting, that we will revisit in the simulation and real-data examples.  If we model the data within a segment as being independent and identically distributed, drawn from a Gaussian distribution with mean $\mu$ and variance $\sigma^2$, then the log-likelihood 
of data $y_{(s+1):t}$, up to a common additive constant, would be 
\[
\ell(y_{(s+1):t};\mu,\sigma)=   -\frac{(t - s)}{2} \log(\sigma^2) - \frac{1}{2\sigma^2} \sum_{j = s + 1}^{t} (y_j - \mu)^2.
\]
For detecting a change in mean, assuming $\sigma^2$ is a known common variance for all observations, we would maximise this log-likelihood with respect to $\mu$. 
The cost associated with a segment could then be minus twice the
maximum of this log-likelihood,

\[
\mathpzc{C}(y_{(s+1):t}) = (t - s) \log(\sigma^2) + \frac{1}{\sigma^2} \sum_{j = s+1}^{t} \left(y_j - \frac{1}{t-s}\sum_{i=s+1}^t y_i\right)^2.
\]
When this cost is summed over segments, the sum of the $(t-s) \log(\sigma^2)$ term is just $n\log(\sigma^2)$ regardless of the segmentation. Hence this term can be dropped from the cost function without
affecting the optimal segmentations as defined by either the constrained or penalised optimisation problems. This segment cost is thus equivalent to using the remaining term on the right-hand side, which
is just a square error cost.

For detecting a change in both mean and variance, calculating the segment cost  would involve using minus twice the log-likelihood after maximising over both $\mu$ and $\sigma$. This gives a segment cost, 
\begin{equation} \label{eq:cmv}
\mathpzc{C}(y_{(s+1):t})=(t - s)\left\{\log\left[\frac{1}{t-s} \sum_{j = s+1}^{t} \left(y_j - \frac{1}{t-s}\sum_{i=s+1}^t y_i\right)^2 \right]+1\right\}.
\end{equation}

\subsection{Finding optimal segmentations} 
Both the constrained and the penalised optimisation problems can be solved by searching the solution space using dynamic programming methods \citep{Bellman1962}. The algorithms in each case
have been called Segment Neighbourhood search and Optimal Partitioning respectively.

\subsubsection{Segment-Neighbourhood} \label{Sec:SN}
\cite{IVANE.AUGERandCHARLESE.LAWRENCE1989} introduced the Segment Neightbourhood (SN) search method which is used to solve the constrained problem in (\ref{eqn:const}).  
This method involves specifying the maximum number of changepoints to allow, $M$, and then calculating the cost of all possible optimal segmentations with 0 to $M$ changepoints. 
The optimal number of changepoints can then be calculated by (\ref{eq:pen_cp}).  The computational cost for this method is $\mathcal{O}(Mn^2)$ and thus this method 
scales poorly when analysing large data sets with a large number of possible changepoints.


\subsubsection{Optimal Partitioning}\label{Sec:OP}
In order to solve the penalised minimisation problem in (\ref{eqn:pen_like}), \cite{Jackson2005} introduced a method also based on dynamic programming: Optimal Partitioning (OP).  Optimal Partitioning is a recursive process which relates the minimum value of (\ref{eqn:pen_like}) to the cost of the optimal segmentation of the data prior to the last changepoint plus the cost of the segment
from the last changepoint to the current time point.  For the data up to time $s$, $y_{1:s}$, we let $\boldsymbol{\tau_s}$ be the set of all possible number and position of changepoints for segmenting
the data: 
$\boldsymbol{\tau_s} = \{ \boldsymbol{\tau} : 0 = \tau_0 < \tau_1 < \cdots < \tau_m < \tau_{m+1} = s\}$.  If we denote the minimisation of (\ref{eqn:pen_like}) for data $y_{1:t}$ by $F(t)=Q(y_{1:t};\beta)$, with $F(0)=0$, then this can be calculated recursively by: 

\begin{align}
F(t) &= \min_{\tau \in \boldsymbol{\tau_t}} \left \{\sum_{i=1}^{m+1} [\mathpzc{C}(y_{(\tau_{i-1}) + 1:\tau_i}) + \beta] \right\}, \\ \nonumber 
&= \min_{s\in\{0,\ldots,t-1\}} \left \{ \min_{\tau \in \boldsymbol{\tau_s}} \sum_{i=1}^{m} [\mathpzc{C}(y_{(\tau_{i-1} + 1):\tau_i}) + \beta ]  + \mathpzc{C}(y_{(s+ 1):t}) + \beta  \right \},\nonumber \\ 
&= \min_{s\in\{0,\ldots,t-1\}} \{ F(s) + \mathpzc{C}(y_{(s+ 1):y}) + \beta  \} \label{eq:OP} .
\end{align}

This recursion can be interpreted as stating that the minimum cost of segmenting $y_{1:t}$ given the last changepoint is at time $s$ is the optimal cost for segmenting data up to time 
$s$ plus the cost of adding a changepoint and the cost for the segment $y_{(s+1):t}$. 
The value of $s$ which attains the minimum of (\ref{eq:OP}) is the position of the last changepoint in the optimal segmentation of $y_{1:t}$.

These recursions are  solved for $t = 1,2,...,n$.  The cost for solving the recursion for time $s$ is linear in $s$, so the overall computational cost is $\mathcal{O}(n^2)$.  Extracting the set of changepoints in the optimal segmentation is achieved by a simple recursion backwards through the data. We first extract the position of the last changepoint in $y_{1:n}$. If this is at
time $\tau$, we then find the next changepoint as it is the last changepoint in the optimal segmentation of $y_{1:\tau}$.  This is repeated until we have a changepoint equal to 0.  

\subsubsection{PELT}\label{Sec:PELT} 

Recently \citet{Killick2012} introduced a modification of Optimal Partitioning; Pruned Exact Linear Time (PELT).  This methods removes values of $\tau$ which can never be minima 
from the minimisation performed at each iteration of the Optimal Partitioning algorithm.  They show that if there exists a constant $K$ such that for all $s<t<T$, 
\begin{align}
\mathpzc{C}(y_{(s+1):t}) + \mathpzc{C}(y_{(t+1):T}) + K \leq \mathpzc{C}(y_{(s+1):T}), \label{eqn:prune2.1}
\end{align}
and for $t > s$, if 
\begin{align}
F(s) + \mathpzc{C}(y_{(s+1):t}) + K \geq F(t), \label{eqn:prune2}
\end{align}
then at a future time $T >t$, $s$ can never be the optimal last changepoint prior to $T$. Therefore $t$ can be removed from the set of possible values of the most recent changepoint that is
searched over in the Optimal Partitioning recursion.  If the cost function is minus the log-likelihood then the constant $K$ in the above function would be 0. 
Pseudo-code for PELT can be found in Algorithm \ref{Alg:PELT}.

\begin{algorithm}
\SetKwData{Left}{left}
\SetKwData{This}{this}
\SetKwData{Up}{up}
\SetKwFunction{Union}{Union}
\SetKwFunction{FindCompress}{FindCompress}
\SetKwInOut{Input}{input}
\SetKwInOut{Output}{output}
\SetKwInOut{Return}{return}
\caption{PELT}
\Input{A data set of the form $y_{1:n} = (y_1,y_2,...,y_n)$\;
\hspace{15 mm} A cost function $\mathpzc{C}(\cdot)$ dependent on the data\;
\hspace{15 mm} A penalty constant $\beta$, and a constant $K$ that satisfies (\ref{eqn:prune2.1}) for all $s<t<T$. 
}
\Output{Details of the optimal segmentation of $y_{1:t}$ for $t=1,\ldots,n$.}
\BlankLine
\emph{Let cp(0) = 0, rescp(0) = 0, $F(0) = 0,~~ m(0) = 0$ and $R_1 = \{0\}$}\;
\For{$t \in \{1,...,n\}$}{
1. Calculate $F(t) = \min_{s \in R_{t}} [F(s) + \mathpzc{C}(y_{(s+1):t}) + \beta]$\;
2. Let $cp(t) = \arg\min_{s \in R_{t}} \{ [F(s) + \mathpzc{C}(y_{(s+1):t}) + \beta]\}$\; 
3. Let $m(t) = m(cp(t))+1$\;
4. Set $rescp(t) = [rescp(cp(t)),cp(t)].$\;
5. Set $R_{t+1} = \{s \in R_{t} : F(s) + \mathpzc{C}(y_{(s+1):t})  < F(t)\}$.
}
\BlankLine
\Return{$rescp(n)$: the changepoints in the optimal segmentation of $y_{1:n}$; \\and for $t=1,\ldots,n$;\\
\hspace*{0.7em} $cp(t)$: the most recent changepoint in the optimal segmentation of $y_{1:t}$;\\
\hspace*{0.7em} $m(t)$: the number of changepoints in the optimal segmentation of $y_{1:t}$;\\
\hspace*{0.7em} $F(t)$: the optimal cost value of the optimal segmentation of $y_{1:t}$.
} 
\BlankLine
\label{Alg:PELT}
\end{algorithm} 

\cite{Killick2012} show that, under certain regularity conditions, the expected computational cost of PELT is $\mathcal{O}(n)$.  
In particular, these regularity conditions require that the number of changepoints increases linearly as the size of the data increases.  
Code implementing this algorithm can be found in the R changepoint package, \cite[]{Rpackage}, with the supporting documentation found in, \cite{Rpackagemanual}. 

%% file: method2.tex
%
%
%
%
\section{Algorithm for a range of penalty values}\label{sec:method}

Algorithms for solving the penalised optimisation problem have been shown to be quicker than those for the constrained problem, however the performance of the penalised approach depends on the penalty value.  
In this section we propose a method which solves the penalised optimisation problem  (\ref{eqn:pen_like}) for a range of penalty values, $\beta$.  
This method finds the optimal segmentations for a different number of segments without incurring as large a computational cost as solving the constrained optimisation problem
for a range of $m$ (the number of changepoints). 

The key to developing an efficient algorithm is to identify those values of $\beta$ for which the penalised optimisation needs to be solved. Ideally for each value of $\beta$ we use we would find a different
optimal segmentation, each corresponding to a different number of changepoints. Whilst we cannot guarantee to achieve this, we can use a relationship between the penalised and constrained optimisation problems
in order to sequentially choose values of $\beta$ in an optimal manner. Furthermore we can re-use calculations from solving the penalised optimisation problem for earlier choices of $\beta$ to speed up the solution
of the penalised optimisation problem for a new value of $\beta$. We develop such an algorithm in the rest of this section. This algorithm can be used within any approach to solving the penalised optimisation problem, but for
concreteness below we assume that PELT is used.
  
\subsection{Link between optimisation problems}
As before, we have $Q_m(y_{1:n})$ as the minimum cost for the constrained optimisation problem (\ref{eqn:const}) and $Q(y_{1:n},\beta)$ as the minimum cost of the penalised optimisation problem (\ref{eqn:pen_like}). 
These
costs can be linked by defining the minimum cost for the penalised optimisation problem subject to the number of changepoints being $m$:
\begin{align}
P_m(\beta) = Q_m(y_{1:n}) + (m+1) \beta. 
\label{eqn:line}
\end{align}
Then we have, for any $\beta$, 
\begin{align} 
Q(y_{1:n}, \beta) = \min_m P_m(\beta). \label{eqn:relationship}
\end{align} 

\begin{figure}[t]
\centering
\includegraphics[width=\linewidth]{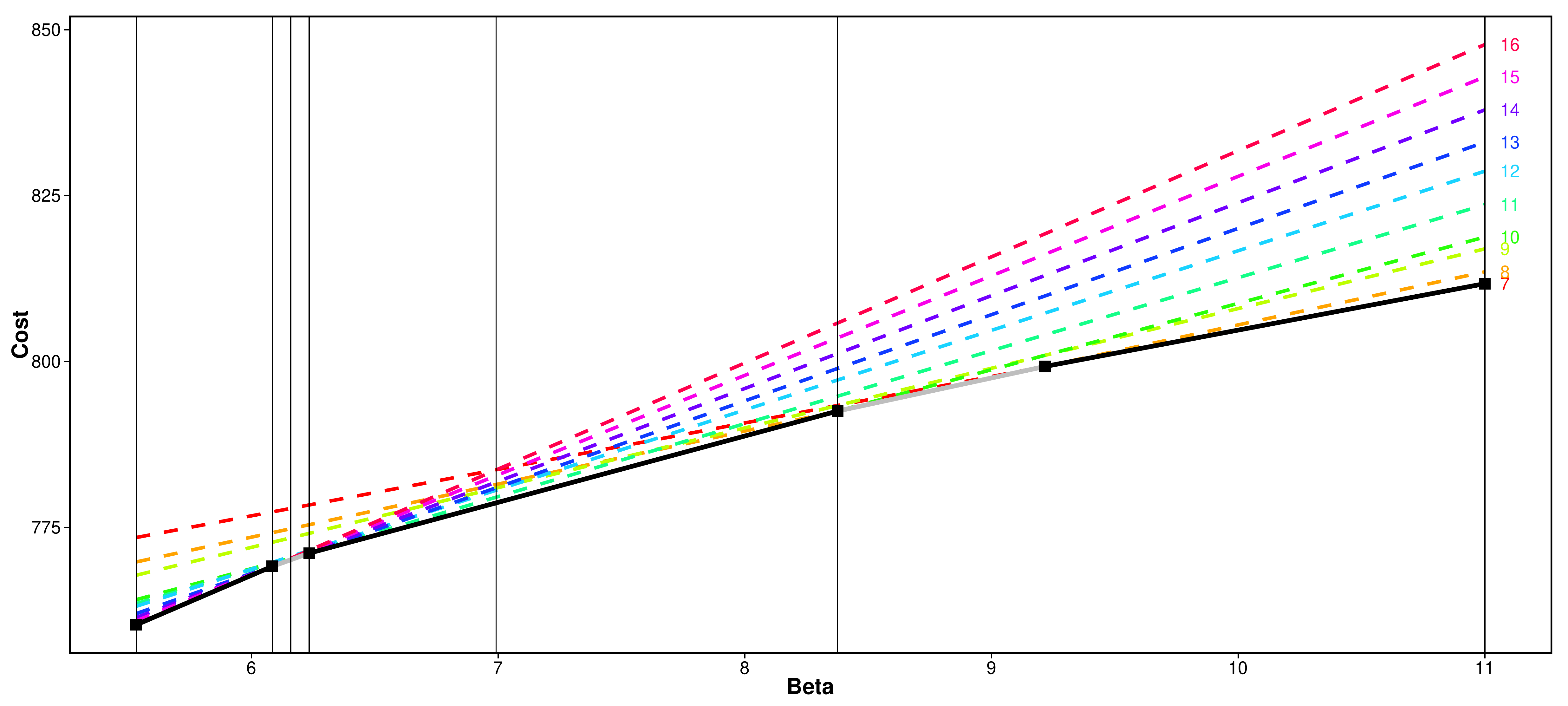} \caption{Graphical representation of the relationship between the constrained and penalised approaches.  
The coloured dashed lines are the costs associated with a different number of changepoints plotted against different penalty terms $\beta$ (\ref{eqn:line}). 
The numbers on the right hand side are the number of changes detected.  The solid dark line shows the optimal value of \emph{$Q(y_{1:n},\beta)$} over the range of $\beta$.  
The solid line is split in to 6 subregions highlighted by different shades and the black squares.  
These indicate the intervals where the optimal number of changepoints is the same for all values of the penalty within the interval.
The set of $\beta$ values for which PELT was run to find all optimal segmentations for $\beta\in[5.5,11]$ are shown by the vertical dotted lines.}
\label{Fig:relationship}
\end{figure}

Figure \ref{Fig:relationship} shows example $P_m(\beta)$ lines, and the corresponding $Q(y_{1:n},\beta)$ curve for a range of penalty values, $\beta \in [5.5,11]$.  
There are a few important points of interest to note from this plot.  Firstly we can clearly see the relationship between the constrained and penalised problems.  
For example it is evident that using a penalty, $\beta = 10 $ and minimising a penalised cost function gives the same optimal segmentation as
solving the constrained optimisation problem with $m = 7$.  
Additionally we can see that as $\beta$ increases the optimal number of changepoints decreases.  By looking at the dashed lines we can see that not all of the
possible number of changes are optimal for some $\beta$. For our example segmentations with  $m = 9, 11, 12, 14$ or $15$ are never optimal choices for any $\beta$.  

Additionally in Figure \ref{Fig:relationship} we can see that the penalty values can be partitioned into intervals which all have the same value of $m$.  
For instance for all $\beta \in [8.2,9.2]$ the resulting $m$ is 8.  This suggests that if we can learn the boundaries of these intervals, we can use that information to solve the penalised optimisation
problem for values of $\beta$ which will correspond to different optimal segmentations. In particular we only needed 
to run PELT for the penalty values indicated on the plot by the vertical lines in order to find all optimal segmentations for $\beta\in[5.5,11]$.  
The next Section describes how we find these values of $\beta$ for which we use in PELT.

\subsection{Theoretical Results}\label{sec:theory}

We now consider the case where we have solved the penalised optimisation problem for two values of penalty, $\beta_0$ and $\beta_1$. The following result describes what we can say about the solutions
to the penalised optimisation problems for other penalty values between $\beta_0$ and $\beta_1$. This result is key to our approach for choosing sequentially the $\beta$ values for which we solve the penalised optimisation problem.\\

Before stating these results we introduce some notation. For any $\beta$ we let $m(\beta)$ be the number of changepoints in the segmentation that is optimal for solving for the penalised optimisation problem
with the penalty being $\beta$. If there is more than one optimal segmentation, we let $m(\beta)$ be the smallest number of changepoints in those optimal segmentations.  Note that, trivially, $m(\beta)$ will be
a decreasing function.

\begin{theorem}\label{th:1} Let $\beta_0<\beta_1$. 
\begin{itemize}
 \item[(1)] If $m(\beta_0) = m(\beta_1)$ then $m(\beta)=m(\beta_0)$ for all  $\beta\in[\beta_0, \beta_1]$.   
 \item[(2)] If $m(\beta_0) = m(\beta_1) + 1$, define 
 \begin{equation} \beta_{int}=\frac{Q_{m(\beta_1)}(y_{1:n}) - Q_{m(\beta_0)}(y_{1:n})}{m(\beta_0) - m(\beta_1)}.\label{eq:Bint}
 \end{equation}
 Then $m(\beta)=m(\beta_0)$ if $\beta \in [\beta_0,\beta_{int})$ and $m(\beta)=m(\beta_1)$ if $\beta \in [\beta_{int},\beta_1]$.
 \item[(3)] If $m(\beta_0) > m(\beta_1) + 1$, and $m(\beta_{int}) = m(\beta_1)$ where $\beta_{int}$ is defined by (\ref{eq:Bint}), 
 then  $m(\beta)=m(\beta_0)$ if $\beta \in [\beta_0,\beta_{int})$ and $m(\beta)=m(\beta_1)$ if $\beta \in [\beta_{int},\beta_1]$.
\end{itemize}
\end{theorem}
\begin{proof}
To simplify notation, write $m_0=m(\beta_0)$ and $m_1=m(\beta_1)$.

Part (1) follows immediately from the fact that $m(\beta)$ is a decreasing function. 

For part (2), note that as $m(\beta)$ is decreasing, then $m(\beta)$ will be equal to either $m_0$ or $m_1$ for all $\beta\in[\beta_0,\beta_1]$. Using (\ref{eqn:relationship}), 
to find the interval of values for which $m(\beta)=m_0$ 
we need to find the values of $\beta$ for which $P_{m_0}(\beta)<P_{m_1}(\beta)$. The value $\beta_{int}$ is just the solution to $P_{m_0}(\beta)=P_{m_1}(\beta)$. This gives the required result.\\

For part (3), first note that as $m(\beta)$ is decreasing, then as $m(\beta_{int})=m_1$ we must have $m(\beta)=m_1$ for all $\beta \in[\beta_{int},\beta_1]$. Thus we only need to show
that for any $m$ with $m_1<m<m_0$ and for all $\beta \in [\beta_0,\beta_{int}]$,
\[
Q_m(y_{1:n})+m\beta\geq Q_{m_0}(y_{1:n})+m_0\beta.
\]

We show this by contradiction. Firstly assume there exists an $m$ with $m_1<m<m_0$ and a $\beta \in [\beta_0,\beta_{int}]$ such that 
\[  
 Q_m(y_{1:n})+m\beta< Q_{m_1}(y_{1:n})+m_0\beta.
\]
As $m<m_0$ and $\beta\leq \beta_{int}$, this implies
\[
 Q_m(y_{1:n})+m\beta_{int}< Q_{m_0}(y_{1:n})+m_0\beta_{int},
 \]
 and by definition of $\beta_{int}$ we then have
 \[
Q_m(y_{1:n})+m\beta_{int}<  Q_{m_1}(y_{1:n})+m_1\beta_{int}.
\]
This then contradicts the condition of part (3) of the theorem, namely that a segmentation with $m_1$ changepoints is optimal for the penalty $\beta_{int}$. 
\end{proof}
\subsection{The Changepoints for a Range of PenaltieS (CROPS) algorithm}    

In the above section we established some key theoretical results which allow us to determine the nature of the resulting number of changepoints when we use penalty values within an interval once we have calculated the results with the end points of the intervals.  We now seek to develop a method to find the number of changepoints using different values of the penalty, $\beta$, in a range $[\beta_{min}, \beta_{max}]$.  Here we introduce the CROPS algorithm, which sequentially calculates the values of $\beta$ which we will use in PELT.   

CROPS begins by first running PELT for penalty values $\beta_{min}$ and $\beta_{max}$. Theorem \ref{th:1} then shows that if we have $m(\beta_{min})=m(\beta_{max})$ or $m(\beta_{min})=m(\beta_{max})+1$ we have found all the
optimal segmentations for $\beta \in [\beta_{min}, \beta_{max}]$. Otherwise we calculate $\beta_{int}$ (\ref{eq:Bint}), the intersection of $P_{m(\beta_{min})}(\beta)$ and $P_{m(\beta_{max})}(\beta)$, then run PELT
with this penalty value. By part (3) of Theorem \ref{th:1} we know that if $m(\beta_{int})=m(\beta_{max})$ then we have found all the optimal segmentations for $\beta \in [\beta_{min}, \beta_{max}]$. Otherwise
we can now consider the intervals $[\beta_{min},\beta_{int}]$ and $[\beta_{int},\beta_{max}]$ separately, and we repeat this procedure on each of those intervals.  
This continues until there are no new intervals to consider. We are able to use the results above to work out the optimal number of changepoints for all penalty values within the interval 
$[\beta_{min}, \beta_{max}]$. 

Pseudo code for this method can be found in Algorithm \ref{Alg:propose}. This code runs a search algorithm such as PELT, say, at all $\beta$ values needed to extract the segmentations that are optimal for some 
$\beta \in [\beta_{min},\beta_{max}]$. If required, the output from these runs can be post-processed to construct the interval of $\beta$ values that each segmentation is optimal for.

 \begin{algorithm}
\SetKwData{Left}{left}
\SetKwData{This}{this}
\SetKwData{Up}{up}
\SetKwFunction{Union}{Union}
\SetKwFunction{FindCompress}{FindCompress}
\SetKwInOut{Input}{input}
\SetKwInOut{Output}{output}
\caption{CROPS algorithm}
\Input{A data set $y_{1:n} = (y_1,y_2,...,y_n)$\;
\hspace{15 mm} Maximum and minimum values of the penalty, $\beta_{min}$ and $\beta_{max}$\;
\hspace{15 mm} {\texttt{CPT}}, an algorithm such as PELT, for solving the penalised optimisation problem.}
\Output{The details of optimal segmentations for each $\beta \in [\beta_{max},\beta_{min}]$.}
\BlankLine
1. Run {\texttt{CPT}} for penalty values $\beta_{min}$ and $\beta_{max}$\; 
2. Set $\beta^* = \{[\beta_{min}, \beta_{max}]\}$\;
\While{$\beta^* \neq \emptyset$}{
3. Choose an element of $\beta^*$; denote this element as $[\beta_0, \beta_1]$\;
\If{$m(\beta_0) > m(\beta_1)+1$}{
4. Calculate $\beta_{int} = \frac{Q_{m(\beta_1)}(y_{1:n}) - Q_{m(\beta_0)}(y_{1:n})}{m(\beta_0) - m(\beta_1)}.$\;
5. Run {\texttt{CPT}} for penalty value $\beta_{int}$\;
6. \If{$m(\beta_{int}) \neq m(\beta_1)$}{
Set $\beta^* = \{\beta^*,[\beta_0,\beta_{int}],[\beta_{int},\beta_1]\}.$\;
}
}
7. Set $\beta^* = \beta^* \setminus [\beta_0, \beta_1]$\;}
\BlankLine
\Return{Output from running CPT for the set of penalty values. 
}
\label{Alg:propose}
\end{algorithm} 

\paragraph{Example} As an example of this algorithm, consider the algorithm for the example shown in Figure \ref{Fig:relationship}.
We initially ran PELT for $\beta_{min} = 5.5$ and $\beta_{max} = 11$.  We found that the number of detected changepoints were 16 and 7 respectively. 
Since the difference is greater than 1 we then calculate $\beta_{int}$, which is the intersection of $P_7(\beta)$ and $P_{16}(\beta)$, and run PELT again with this value.  
This time we find that there are 10 changes at $\beta_{int} = 7$. We then repeat this procedure for the intervals $[5.5,7]$ and $[7,11]$. In each
case we find the corresponding $\beta_{int}$ values,
6.2 and 8.2, and run PELT for these values of the penalty. These produce segmentations with 13 and 8 changepoints respectively. By Theorem \ref{th:1} part (2) we know that
we have found all segmentations for $\beta \in [8.2,11]$. We then repeat this procedure and continue in a similar manner until we have found solutions for all of the intervals.  

\subsection{The Number of Changepoints that are Optimal for some $\beta$}

For the example in Figure \ref{Fig:relationship} we saw some of the optimal segmentations for specific numbers of changepoints
would never be optimal regardless of the penalty value used. Thus using this method will not necessarily 
get the resulting segmentations for all numbers of changepoints, something which you get when you use segment neighbourhood search. 

\begin{figure}[H]
\centering
\includegraphics[width=\linewidth]{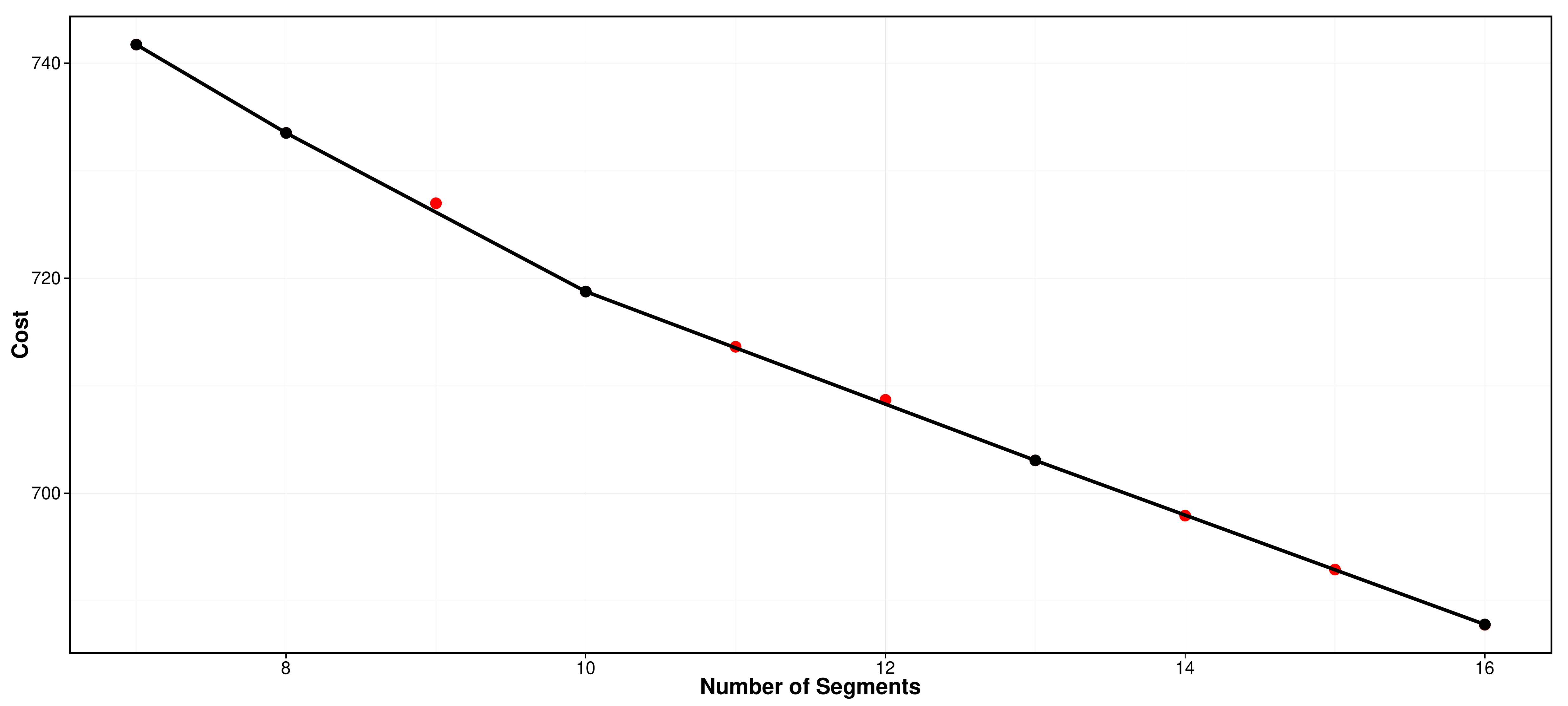} \caption{Cost for the segmentations against the number of changepoints.  
The black dots are the points corresponding to optimal segmentations found by solving the penalised optimisation problem over some range of $\beta$.
The red dots correspond to optimal segmentations other numbers of changepoints.\label{Fig:convex}}
\end{figure}

\cite{Lavielle2005} gives a condition under which a segmentation with $m$ changepoints will be the optimal segmentation for some $\beta$.
Assume that segmentations with $m_1<\cdots<m_k$ changes, for some $k>1$, are optimal as we vary $\beta \in [\beta_{min},\beta_{max}]$. Let $Q_i=Q_{m_i}(y_{1:n})$, for $i=1,\ldots,k$, be the
associated un-penalised cost of these segmentations. We can construct a piece-wise linear line by joining $(m_i,Q_i)$ with $(m_{i+1},Q_{i+1})$ for $i=1,\ldots,k-1$. All values of changepoints, $m$, 
with $m_1<m<m_k$ and for which there is no optimal segmentation will lie above this line. An example is shown in Figure \ref{Fig:convex}.  

One way of expressing this condition is that we will not obtain segmentations for which the average reduction in cost of adding some number of changepoints is more than the average increase in cost of 
removing some number of changepoints. Consider the example in Figure \ref{Fig:convex}. By solving the penalised optimisation problem for a range of $\beta$ we do not
find an optimal segmentation with 9 changepoints. This is because the reduction in cost of going from 8 to 9 changepoints is less than for going from 9 to 10 changepoints. It is hard to construct a criteria under
which the segmentations not found by solving the penalised optimiation problem would be optimal. In fact \cite{Killick2012} show that any segmentation that is optimal under (\ref{eq:pen_cp}) where the penalty function
for adding changepoints, $f(m)$, is concave will be the solution to the penalised optimisation problem for some $\beta$. 
 
 \subsection{Computational Cost}\label{sec:cost}

We now bound the computational cost of our proposed approach. We do this in terms of the maximum number of times PELT would need to be run. The following theorem shows that this is at
most  $m(\beta_{min}) - m(\beta_{max}) + 2$ times.  
 
\begin{theorem}\label{theorem:cost}

 \begin{itemize}
 \item[(1)] If $m(\beta_{0}) = m(\beta_{1})$ then the maximum number of times that PELT is required to be run to find all the optimal segmentations for $\beta\in[\beta_{0},\beta_{1}]$ is $m(\beta_{0}) - m(\beta_{1}) + 2$.
 \item[(2)] If $m(\beta_{0})>m(\beta_{1})$ then the number of times that PELT is required to be run to find all the optimal segmentations for $\beta\in[\beta_{0},\beta_{1}]$ is bounded above by
 \[
  m(\beta_{0}) - m(\beta_{1}) + 1.
 \]
 \end{itemize}
\end{theorem}

\begin{proof}
The proof for part (1) is trivial since we need to run PELT twice, using both $\beta_{0}$ and $\beta_{1}$.  

For the proof of part (2) define $N(m_0, m_1)$ as the maximum (over data sets) of the number of further runs of PELT needed to find all the optimal segmentations in an interval of $\beta$, given we have
run PELT at the lower and upper endpoints of the interval and these have produced segmentations with $m_0$ and $m_1$ changepoints respectively.  
As we have run PELT twice, to prove the theorem we need to show that 
\begin{align}
N(m_0, m_1) \leq m_0 - m_1 - 1. \label{eqn:cost_proof}
\end{align}
Firstly, if  $m_0 - m_1 = 1$ then $N(m_0, m_1) = 0$, which satisfies (\ref{eqn:cost_proof}).  

Now we proceed by induction. For an integer $l>1$ assume that if $m_0 - m_1 < l$ then (\ref{eqn:cost_proof}) holds. 
We need to show that this implies that (\ref{eqn:cost_proof}) holds for $m_0 - m_1=l$.

In this case our first step is to run PELT at the intersection, $\beta_{int}$.  In the worst case scenario we find that $m(\beta_{int}) \neq m_1$ (and hence  $m(\beta_{int}) \neq m_0$ as segmentations with $m_0$ and $m_1$ 
changepoints have the same penalised cost for penalty value $\beta_{int}$).
We then need to consider the sub-intervals below and above $\beta_{int}$ separately.  Since $m(\beta)$ decreases as $\beta$ increases $m_0-m(\beta_{int}) < l$ and $m(\beta_{int})-m_1 < l$. Therefore 
\begin{align}
N(m_0, m_1) &= 1 + N(m_0, m(\beta_{int})) + N(m(\beta_{int}), m_1) \nonumber\\
&\leq 1 + [m_0-m(\beta_{int})- 1] + [m(\beta_{int}) - m_1 - 1]\nonumber \\
&= 1 + m_0 - m_1 - 2 \nonumber\\ 
&= m_0 - m_1 - 1. \nonumber
\end{align}
which satisfies (\ref{eqn:cost_proof}) as required.  
\end{proof} 

\subsubsection{Recycling Calculations}
It is further possible to speed up Algorithm \ref{Alg:propose} by recycling some of the calculations from different runs of PELT.  
In Algorithm \ref{Alg:PELT} we calculate and store the minimum penalised cost for segmenting data $y_{1:t}$, the number
of changepoints in this segmentation for $t=1,\ldots,n$ and the position of the most recent changepoint up to time $t$. 
If PELT was run with penalty value $\beta$ we denote these values as $F(t,\beta)$, $m(t,\beta)$ and $cp(t,\beta)$ respectively.
We can re-use these values from previous runs of PELT to precalculate many of the values for a new run. 

Assume we have run PELT with penalty values $\beta_0$ and $\beta_1$, and are now wanting to run PELT for $\beta_{int}$ where $\beta_0<\beta_{int}<\beta_1$.
Before running PELT for the new value we iterate for $t = 1,...,n$: 
\begin{enumerate}
\item If $m(t,\beta_{0}) = m(t,\beta_{1})$ then set $m(t,\beta_{int}) = m(t,\beta_{0}), ~ cp(t,\beta_{int}) = cp(t,\beta_{0}) $ and 
\begin{align*}
F(t,\beta_{int}) = F(t,\beta_{0}) + m(t,\beta_{int})(\beta_{int} - \beta_{0}).
\end{align*}

\item If $m(t,\beta_{0}) = m(t,\beta_{1}) + 1$ then calculate 
\begin{align*}
a = F(t,\beta_{0}) + m(t,\beta_{0})(\beta_{int} - \beta_{0}),
\end{align*}
and 
\begin{align*}
b = F(t,\beta_{1}) + m(t,\beta_{1})(\beta_{int} - \beta_{1}).
\end{align*}
If $a < b$ then $m(t,\beta_{int}) = m(t,\beta_{0}), ~ cp(t,\beta_{int}) = cp(t,\beta_{0})$ and $F(t,\beta_{int}) = a$; otherwise $m(t,\beta_{int}) = m(t,\beta_{1}), ~ cp(t,\beta_{int}) = cp(t,\beta_{1}) $ and $F(t,\beta) = b$.  
 \end{enumerate}  
We then just need to run PELT to calculate the values of $F(t,\beta_{int}), ~ m(t,\beta_{int})$ and $cp(t,\beta_{int})$  for times $t$ that we have not been able to precalculate them.  


%% file: simulation3.tex
%
%
%



\section{Simulation Study}\label{sec:sims}

This section aims to illustrate why our proposed method is useful in practice.  We firstly show that using CROPS with PELT can be substantially quicker than using Segment Neighbourhood search
to find optimal segmentations for a range of numbers of changepoints. We are also able to use CROPS to efficiently study and compare some different proposals for the choice of the penalty.
Whilst some of these work well when we use the correct model for the data, 
we show that they can give misleading results when the model is mis-specified, something that is likely
to be a feature of real-life applications of changepoint detection. 

\subsection{Simulation Set-Up} 
For our simulation study we consider detecting multiple changes in the mean and variance. We set the cost function to be (\ref{eq:cmv}), which is based on modelling data in each
segment as independent and identically distributed from a Gaussian distribution with unknown mean and variance.

We use two settings for simulating the data to analyse. In the first we simulate data in each segment as independent realisations from a Gaussian distribution, and let the mean and variance of
this distribution vary across segments. This corresponds to the data being simulated from the model we use for analysing the data. The second setting corresponds to a mis-specified model, where we let
the mean of each data point vary slowly with the position within a segment.  

For the mis-specified model, for a segment $k$ we simulate segment standard deviations, $\sigma_k^2$, and an initial mean value, $\mu_k$. If $Y_t$ is in segment $k$ then we simulate our data from
\begin{align}
Y_t \sim N(\nu_t, \sigma_k^2),
\end{align} 
where $\nu_t = \mu_k$ if $t$ is the first point in a segment and $\nu_{t+1} = \nu_{t} + \epsilon_{t}$, $\epsilon_{t} \sim N(0,0.1)$ otherwise.  
For both models we simulate data sets with varying lengths with changepoints distributed uniformly in time but with the constraint that there are at least 20 observations between changepoints.
For a given value of $n$ we simulate data sets with (i) a fixed number of changepoints, $m = 10$, (ii) the number of changepoints increases sub-linearly with $n$, $m = \sqrt{n}/4$, and (iii) the number of changepoints 
increases linearly with $n$, $m = n/100$. 

For both models we generate the (initial) segment means from a Normal distribution with mean 0 and standard deviation 2.5 the segment standard deviations from a Log-Normal 
distribution with mean 0 and standard deviation $\frac{\log(10)}{2}$. 

\subsection{CPU cost}

Firstly we compare the computational cost for running CROPS with PELT to find different segmentations using the penalised minimisation problem in comparison to using Segment
Neighbourhood. Additionally we investigate the improvement the recycling of calculations as suggested in Section \ref{sec:cost} makes. For this simulation we set $\beta_{min}$ = 14 and  $\beta_{max}$ = 40.   
Note in Segment Neighbourhood we set the maximum number of changepoints to be the number of 
changepoints detected using the smallest value of the penalty value.

\begin{figure}
\centering
\begin{subfigure}{0.3\linewidth}
\includegraphics[width=\linewidth]{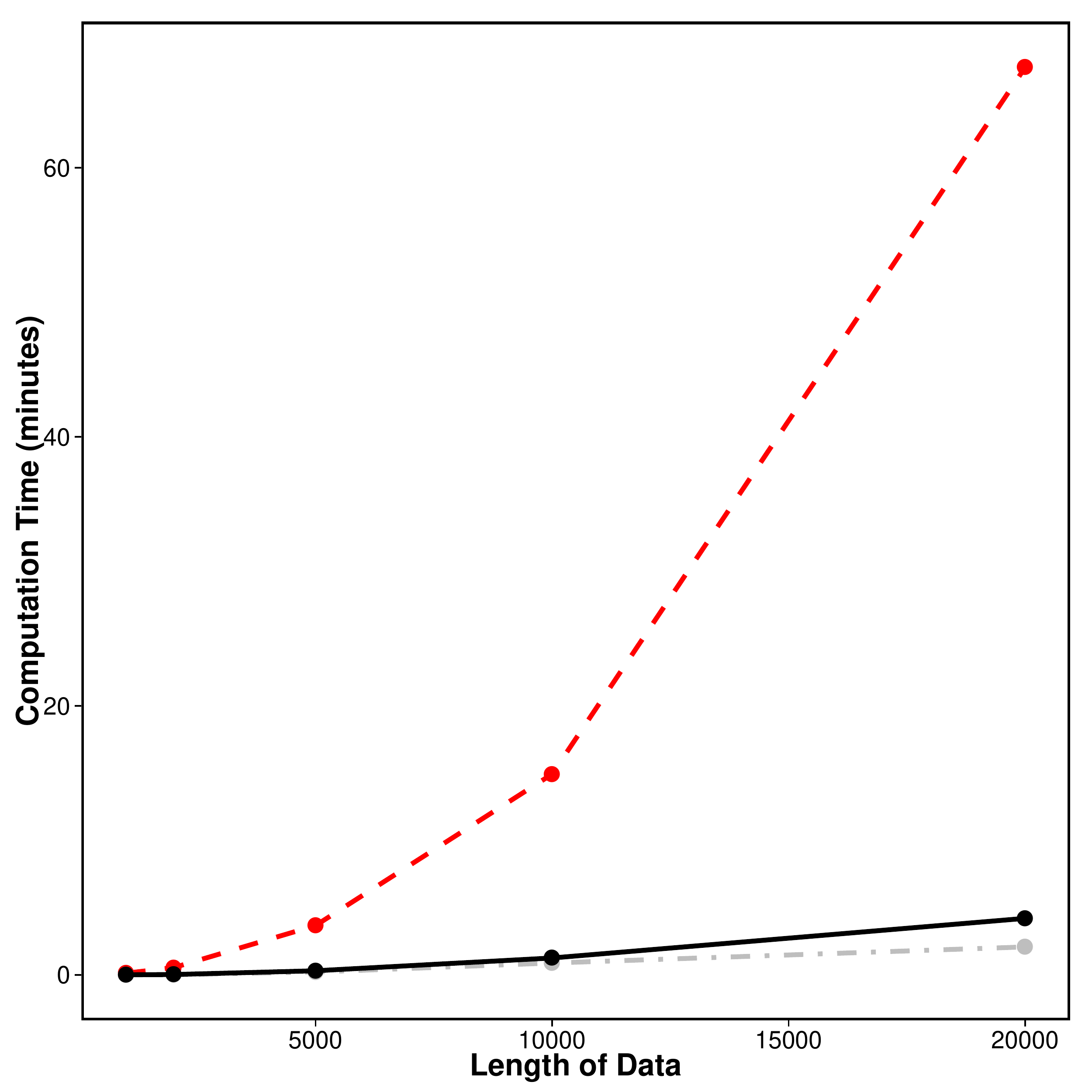} 
\end{subfigure}
~
\begin{subfigure}{0.3\linewidth}
\includegraphics[width=\linewidth]{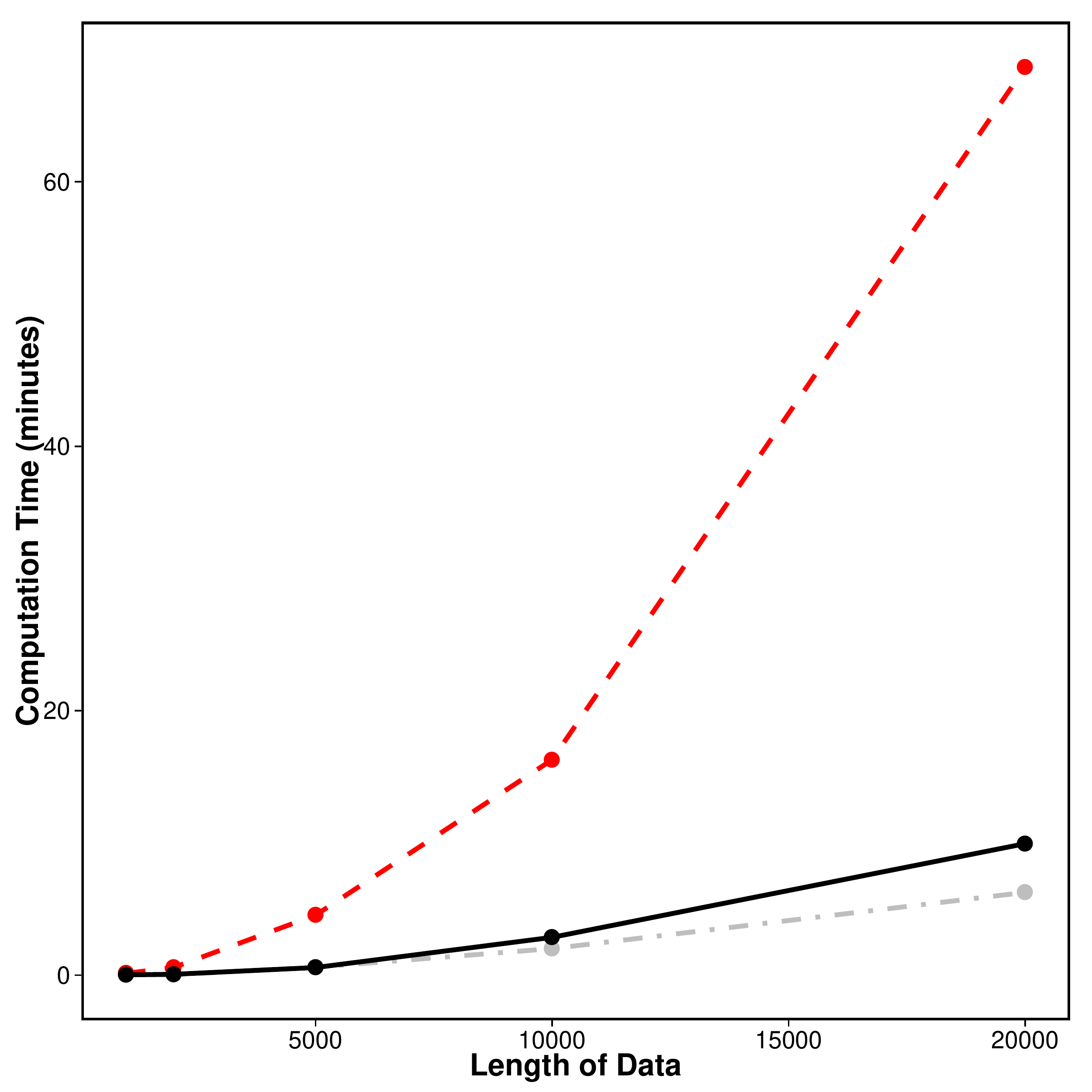}
\end{subfigure}
~
\begin{subfigure}{0.3\linewidth}
\includegraphics[width=\linewidth]{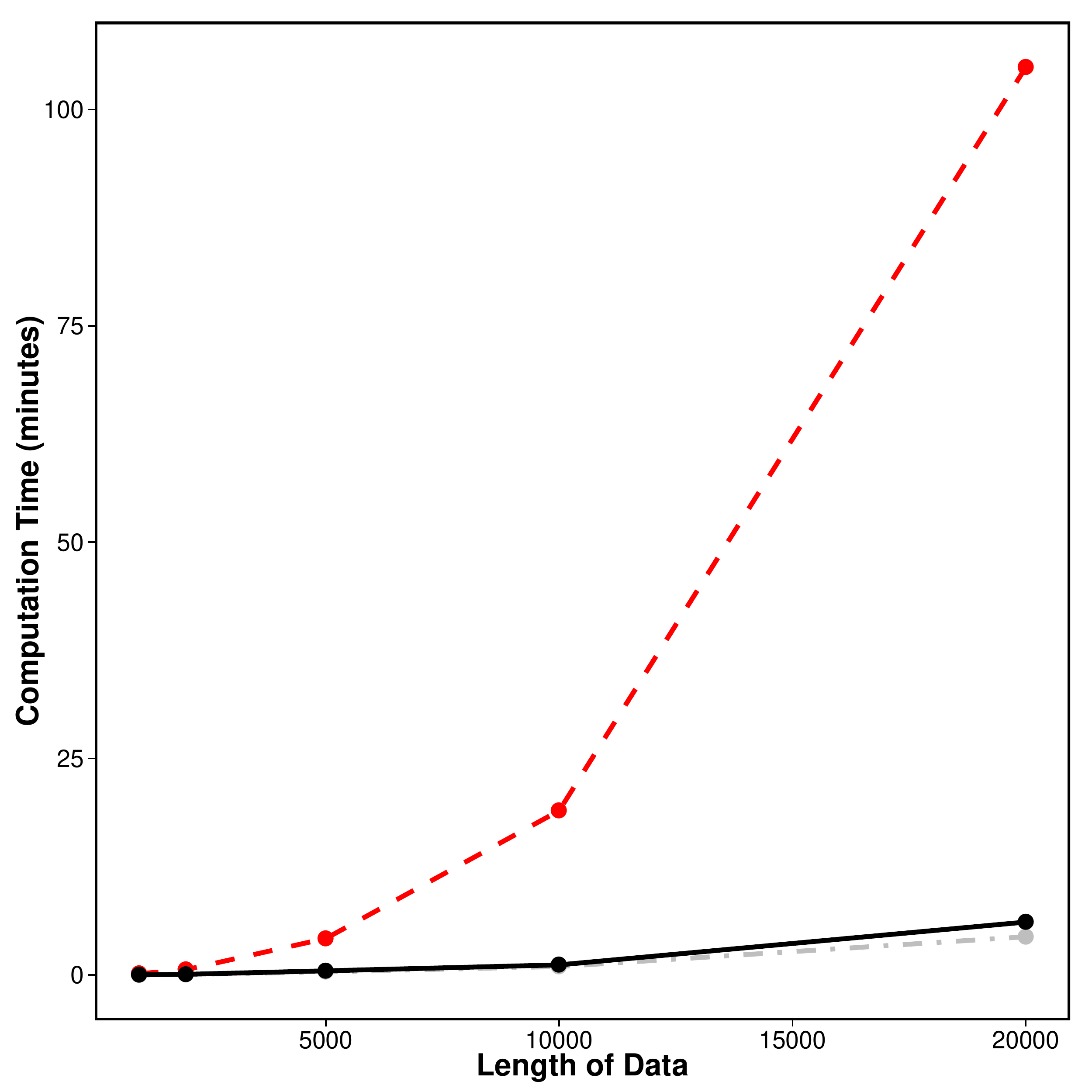}
\end{subfigure}

\begin{subfigure}{0.3\linewidth}
\includegraphics[width=\linewidth]{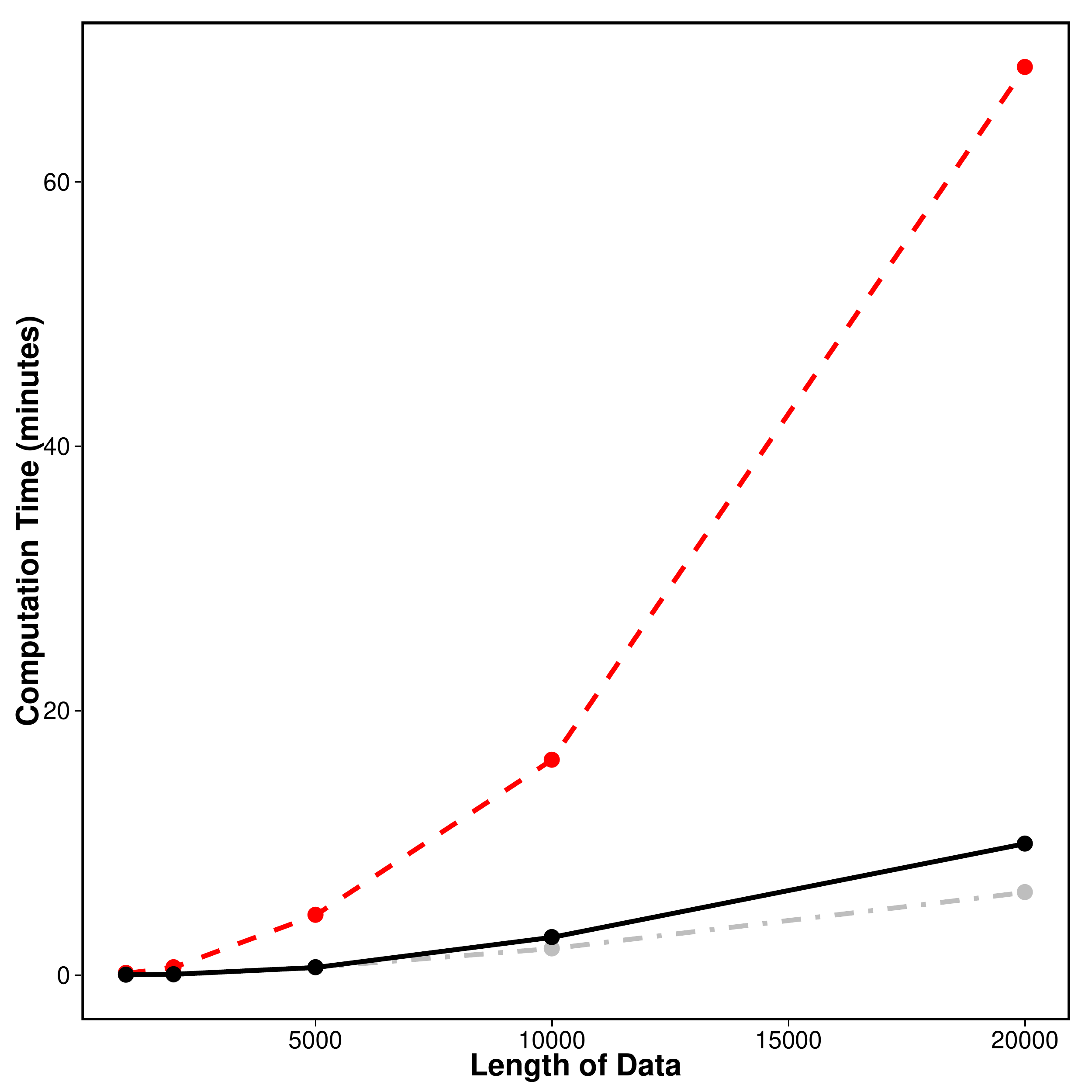} 
\end{subfigure}
~
\begin{subfigure}{0.3\linewidth}
\includegraphics[width=\linewidth]{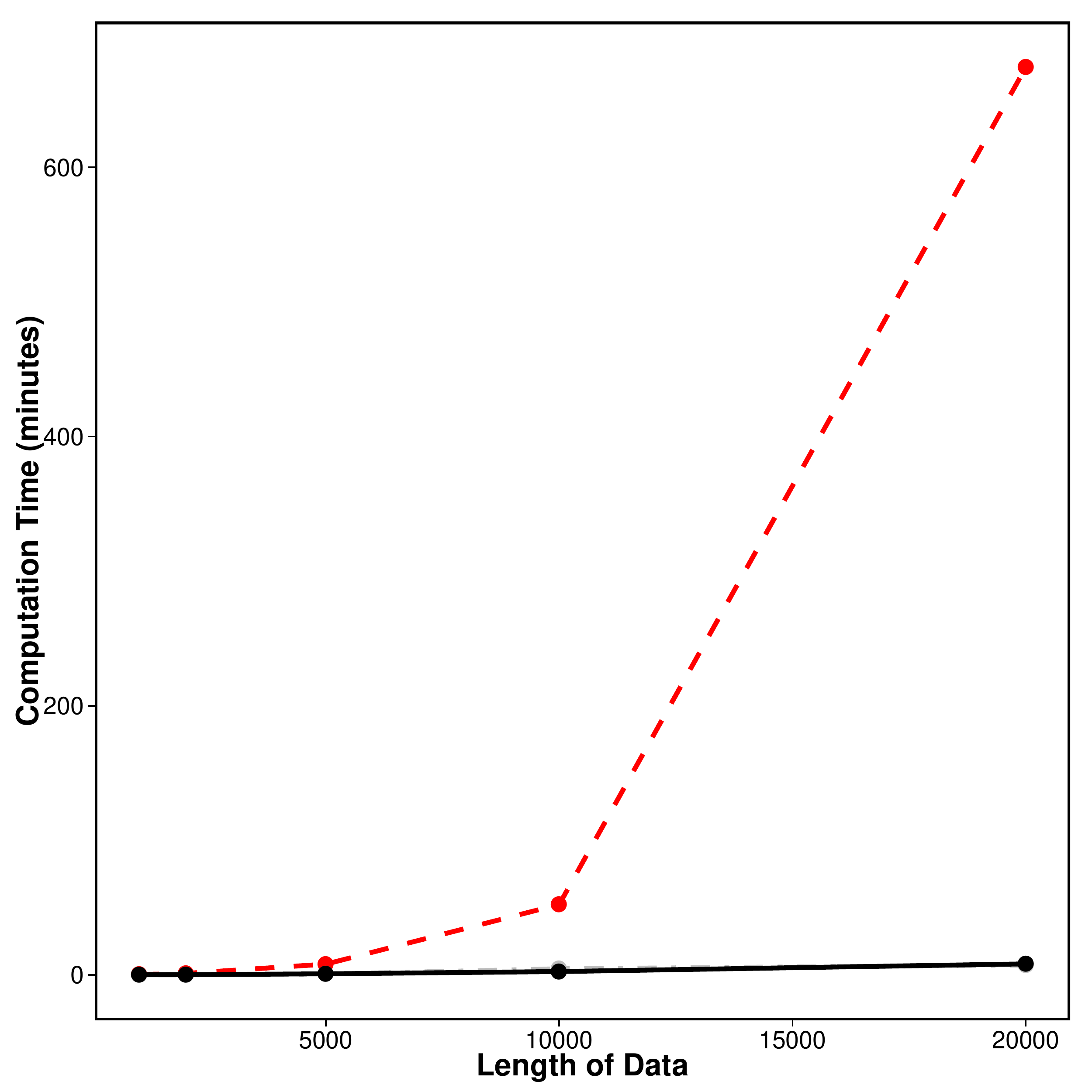}
\end{subfigure}
~
\begin{subfigure}{0.3\linewidth}
\includegraphics[width=\linewidth]{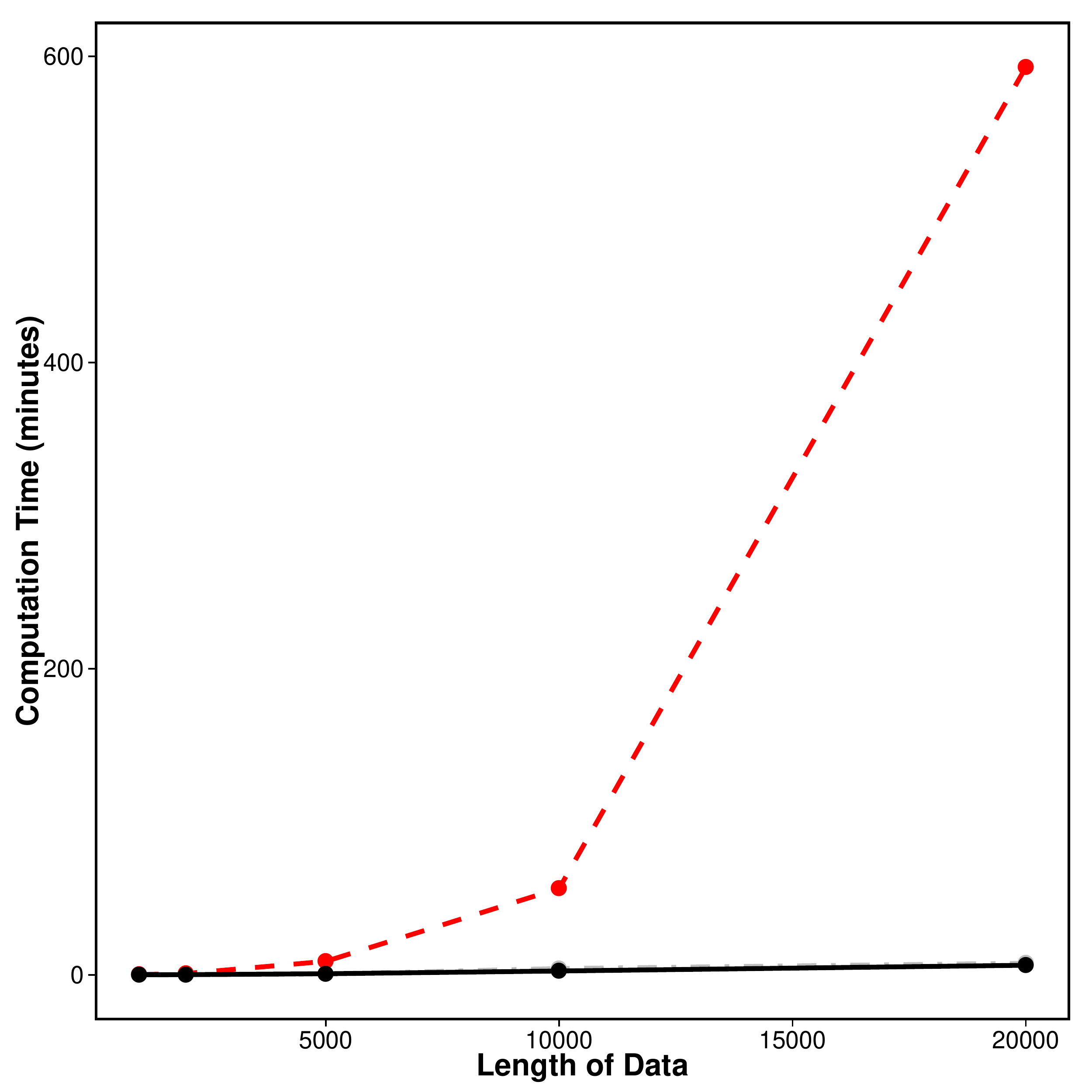}
\end{subfigure}

\caption{CPU cost of using (black, solid) PELT with a range of $\beta$'s to find a range of segmentations, 
(grey, dot-dashed) PELT including the speed-up of Section \ref{sec:cost} and (red, dashed) Segment Neighbourhood.  
Top: true model and bottom: mis-specified model. Left: fixed changepoints ($m=10$), middle: sublinear number of changepoints ($m = \sqrt{n}/4$) and right: linear number of changepoints ($m=n/100$).}
\label{fig:CPU}
\end{figure}

The results for both the Normal model and the misspecified model can be found in Figure \ref{fig:CPU}.  It is evident that solving the penalised optimisation problem using PELT, with and without the speed up improvement, 
for a range of penalty values is substantially quicker than running Segment Neighbourhood. The speed-up appears to increase with data size, and for $n=20,000$ 
the speed-ups were by factors of between 10 and 100.   
The computational cost to run PELT with and without the recycling of the calculations are very similar, with, in general, recycling leading to modest gains in speed.

%


\subsection{Evaluating the Choice of Penalty}\label{sec:optimalb}
For both the true model and the mis-specified model we can use our approach to efficiently evaluate the accuracy of segmentations using different penalty terms such as Schwarz's Information Criterion (SIC), 
Akaike's Information Critrerion (AIC) and Hannan-Quinn.  
We compare accuracy in terms of estimating the number of changepoints, detecting the position of the changepoints, and estimating the segment parameters. 

For the first of these we calculate the range of $\beta$ values that give correct estimates of the number of changepoints. For a given simulation scenario we calculate the average of this range over 100 simulated
data sets, and compare this average with the different penalty choices. To evaluate the accuracy of the estimated changepoint positions, we define an actual changepoint as detected if we infer a changpoint within 10 time points
of its actual location. We call these true positives. We can then define the number of false-positives in a segmentation as the number of inferred changepoints minus the number of actual changepoints detected. We
measure accuracy by looking at the average proportion of changepoints detected, defined as the number of true positives divided by the true number of changepoints; against the average proportion of false positives, the number of false positives divided by the
number of changepoints detected. 

Finally we use a mean square error criteria to evaluate the accuracy of estimates of the segment parameters. For a given segmentation we can calculate the maximum likelihood estimates of the segment mean and standard deviation. Then
for each time-point we compare the estimated parameter values for the segment we infer that time-point belonging to, to the actual parameter values of the segment the time-point is in. We do this separately for the mean
and standard deviation. So if $\hat{\theta_i}$ is the estimated parameter, for example mean, of the observation at time $i$, and $\theta_i$ the true parameter then: 
\begin{align}
\mbox{MSE} = \frac{\sum_{i=1}^{n}(\hat{\theta}_i - \theta_i)^2}{n}.
\end{align}

\begin{figure}
\centering
\begin{subfigure}{0.3\linewidth}
\caption{}\includegraphics[width=\linewidth, height = 5cm]{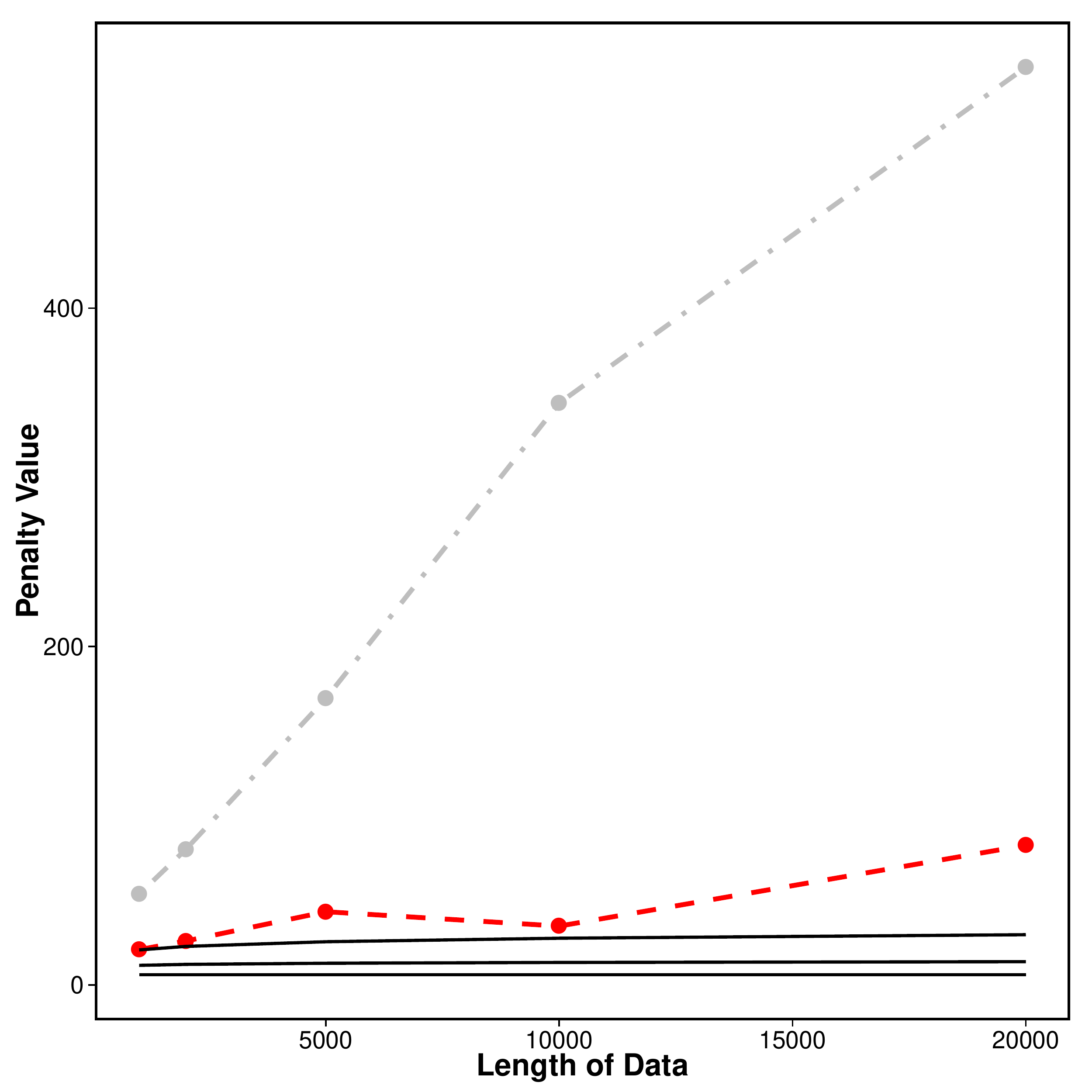} \label{fig:meanvar_res}
\end{subfigure}
~
\begin{subfigure}{0.3\linewidth}
\caption{}\includegraphics[width=\linewidth, height = 5cm]{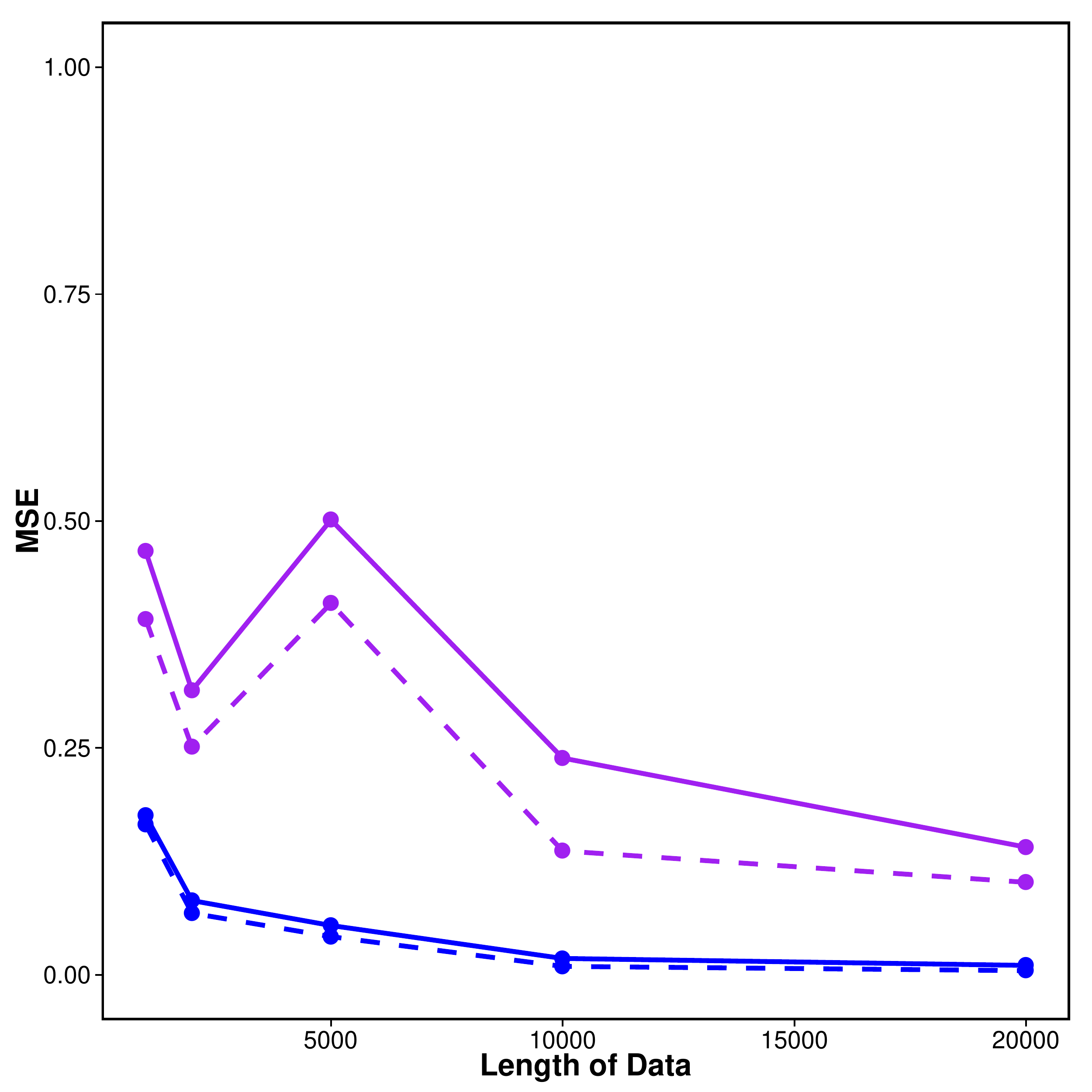}
\label{fig:meanvar_MSE}
\end{subfigure}
~
\begin{subfigure}{0.3\linewidth}
\caption{}\includegraphics[width=\linewidth, height = 5cm]{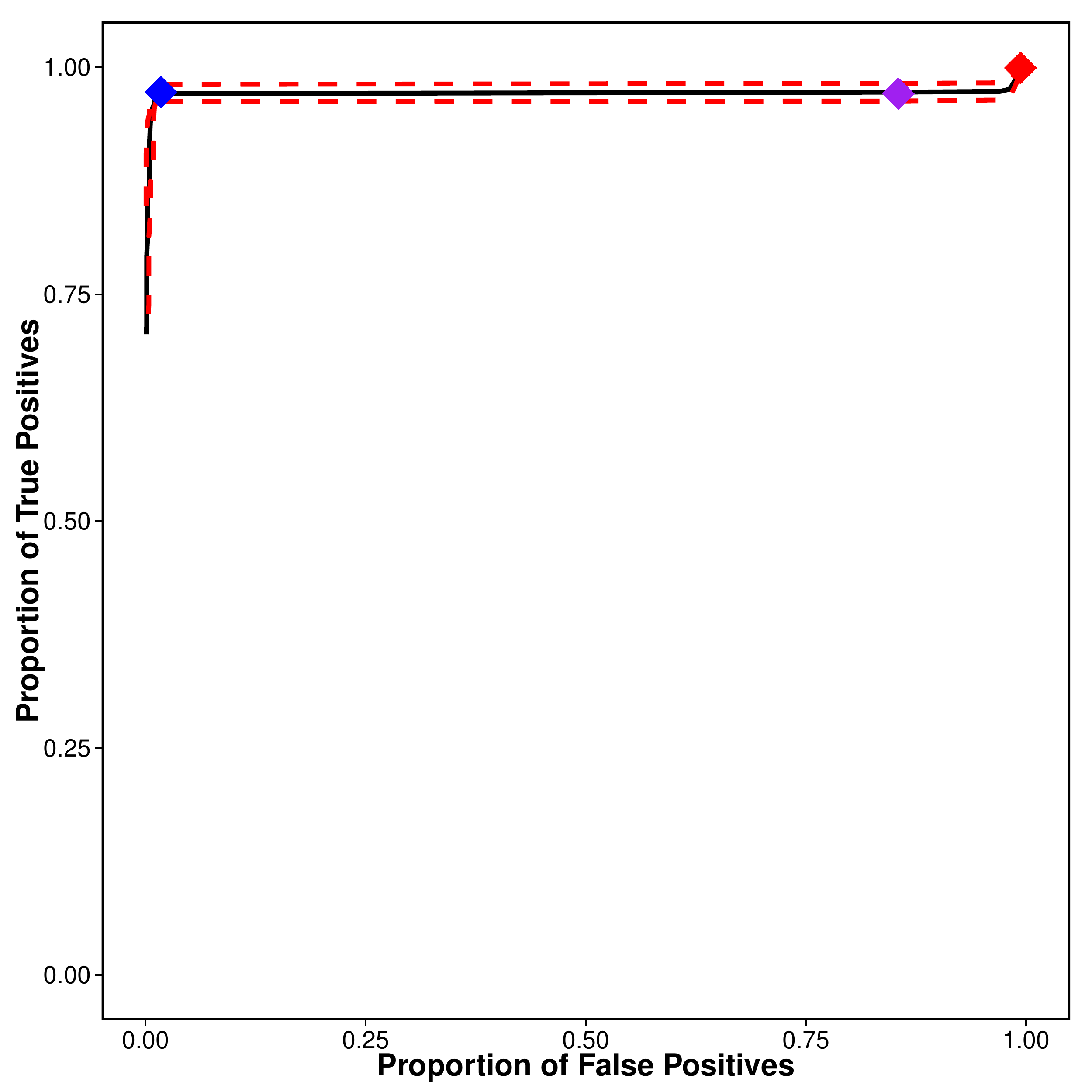}
\label{fig:meanvar_ROC}
\end{subfigure}

\begin{subfigure}{0.3\linewidth}
\includegraphics[width=\linewidth, height = 5cm]{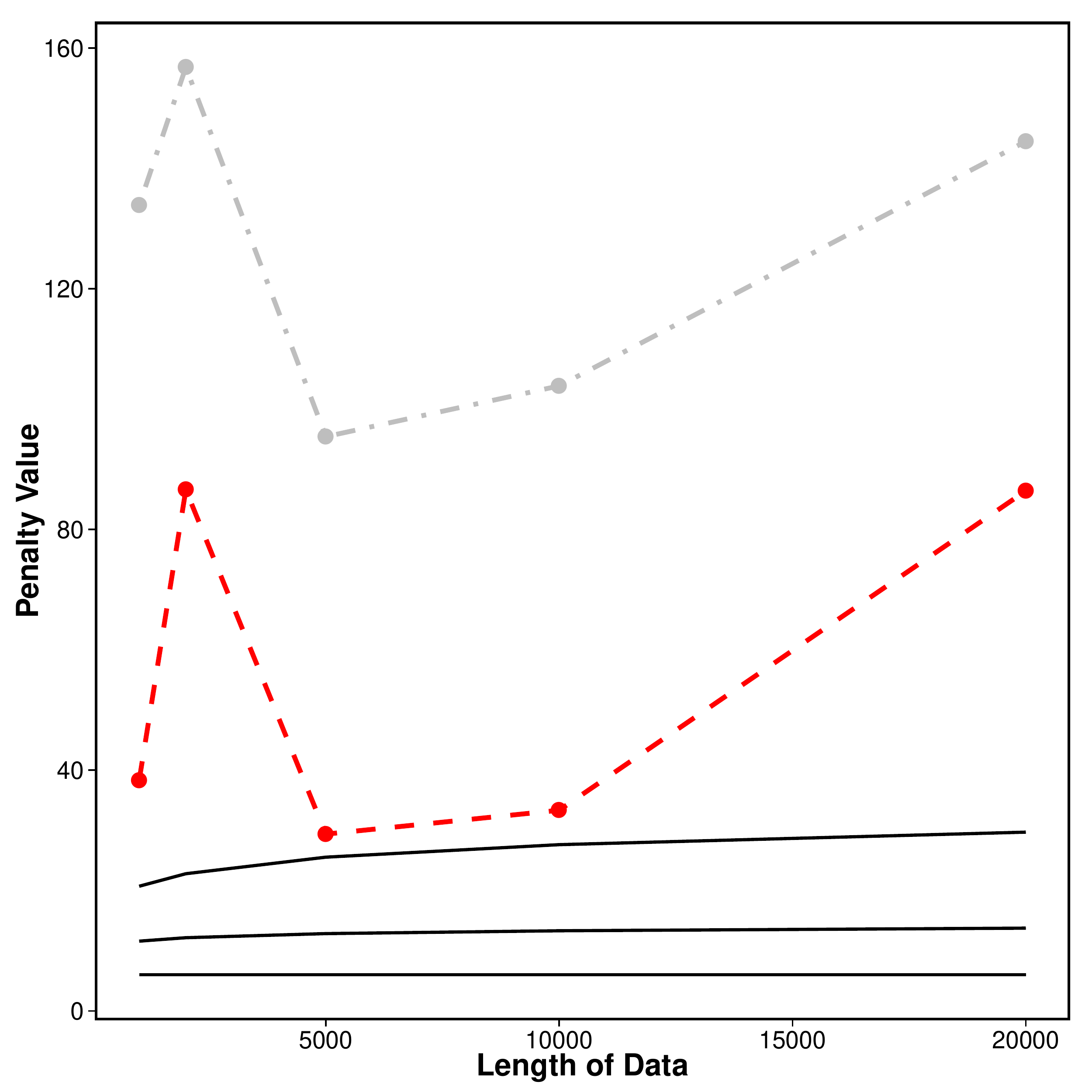}
\end{subfigure}
~
\begin{subfigure}{0.3\linewidth}
\includegraphics[width=\linewidth, height = 5cm]{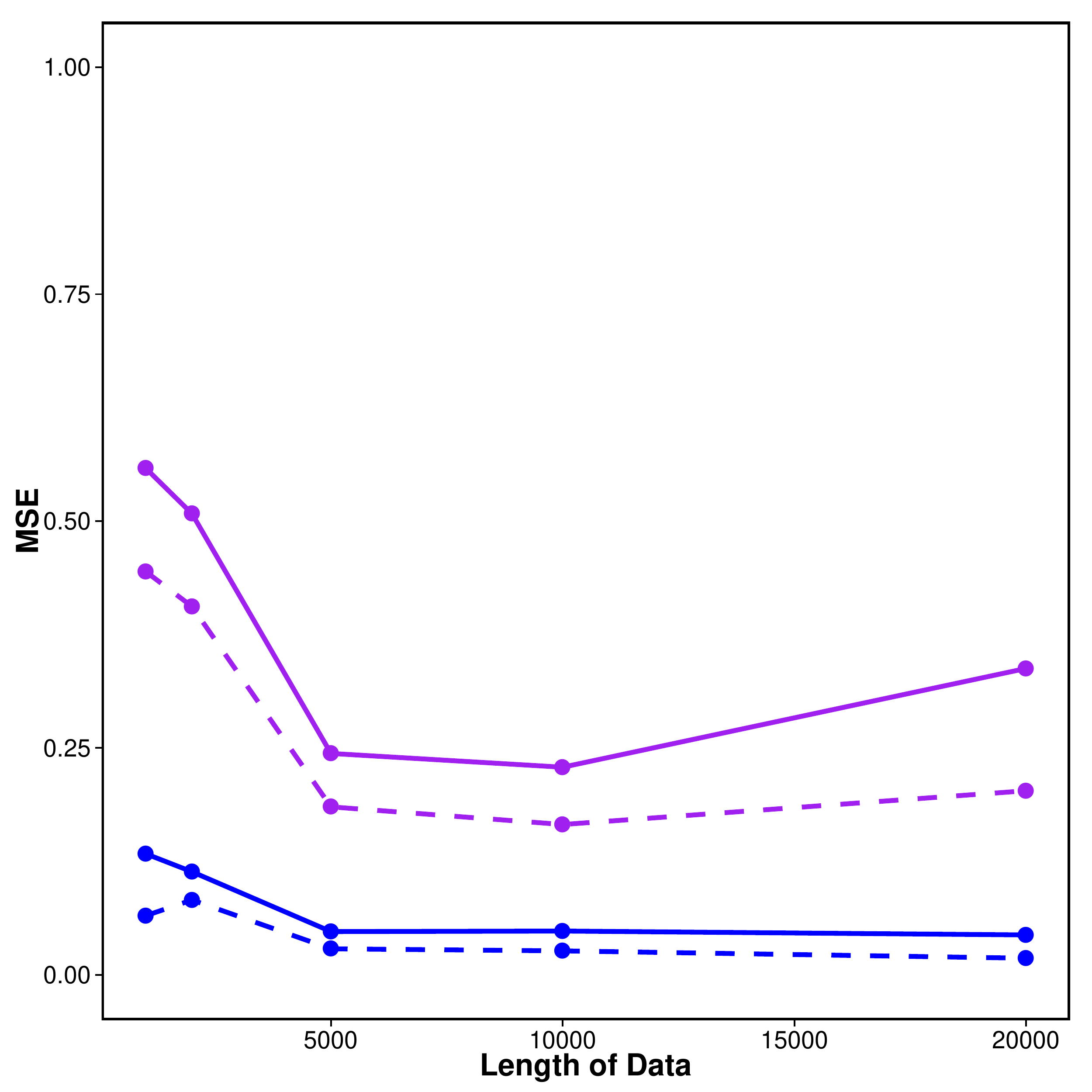}
\end{subfigure}
~
\begin{subfigure}{0.3\linewidth}
\includegraphics[width=\linewidth, height = 5.5cm]{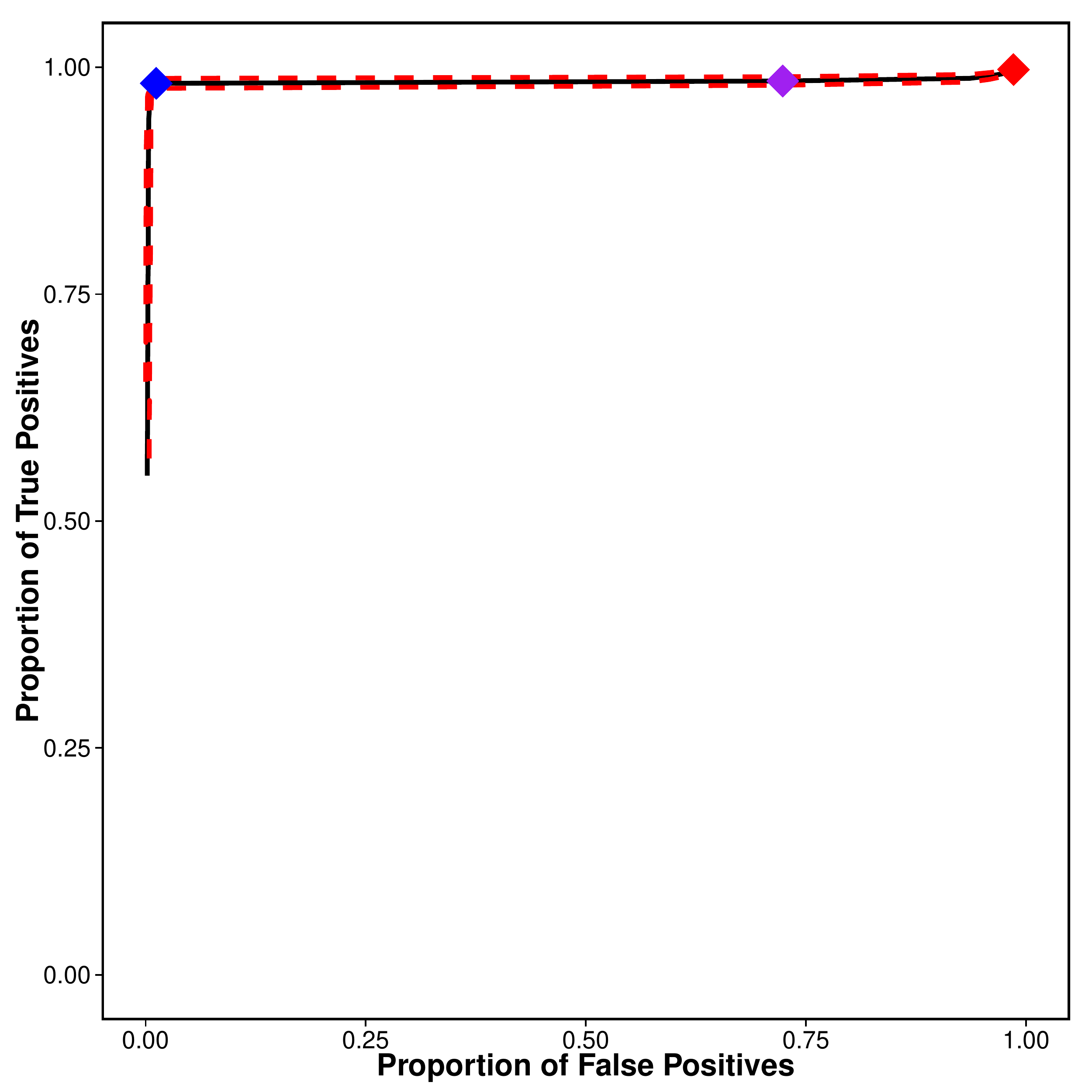}
\end{subfigure}

\begin{subfigure}{0.3\linewidth}
\includegraphics[width=\linewidth, height = 5cm]{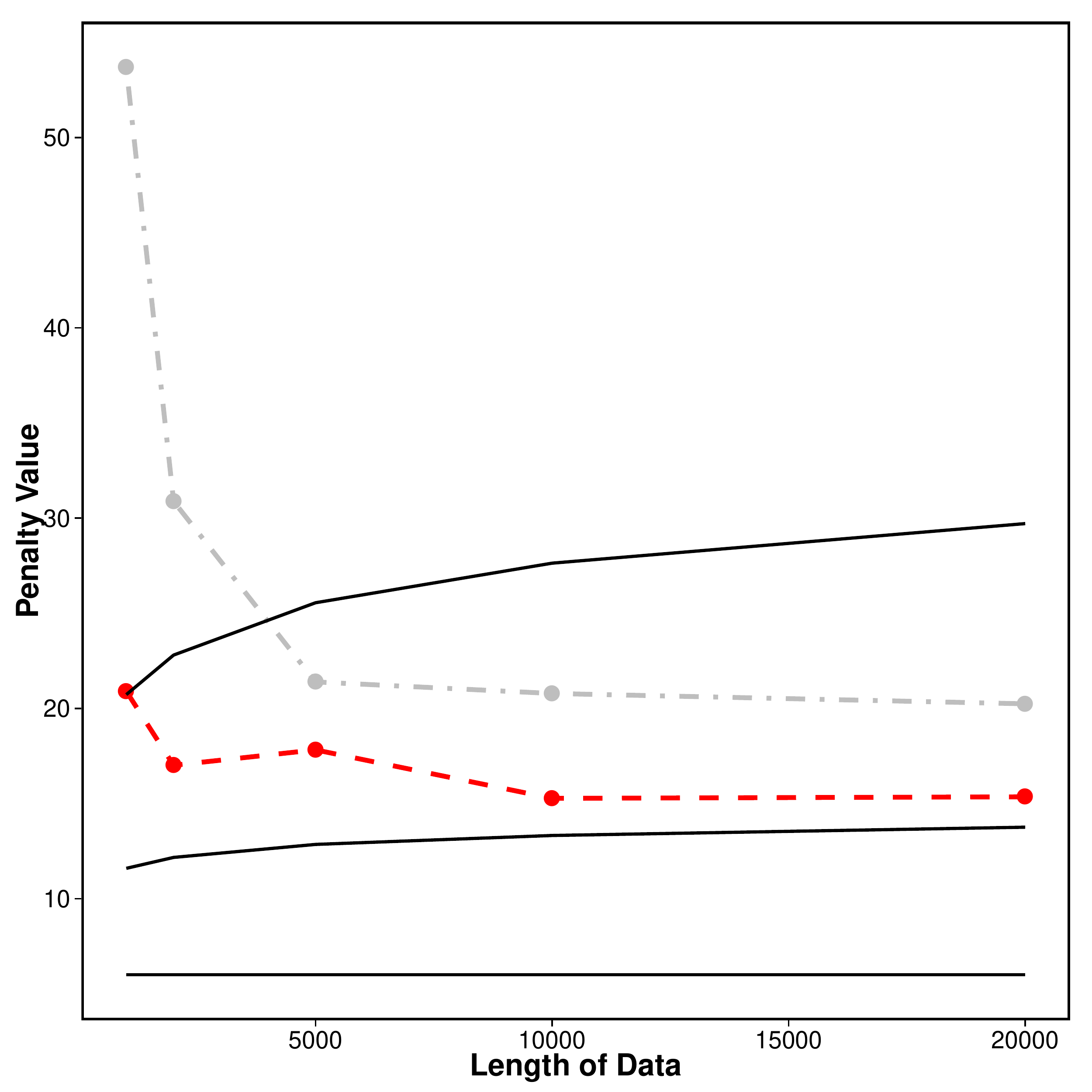}
\end{subfigure}
~
\begin{subfigure}{0.3\linewidth}
\includegraphics[width=\linewidth, height = 5cm]{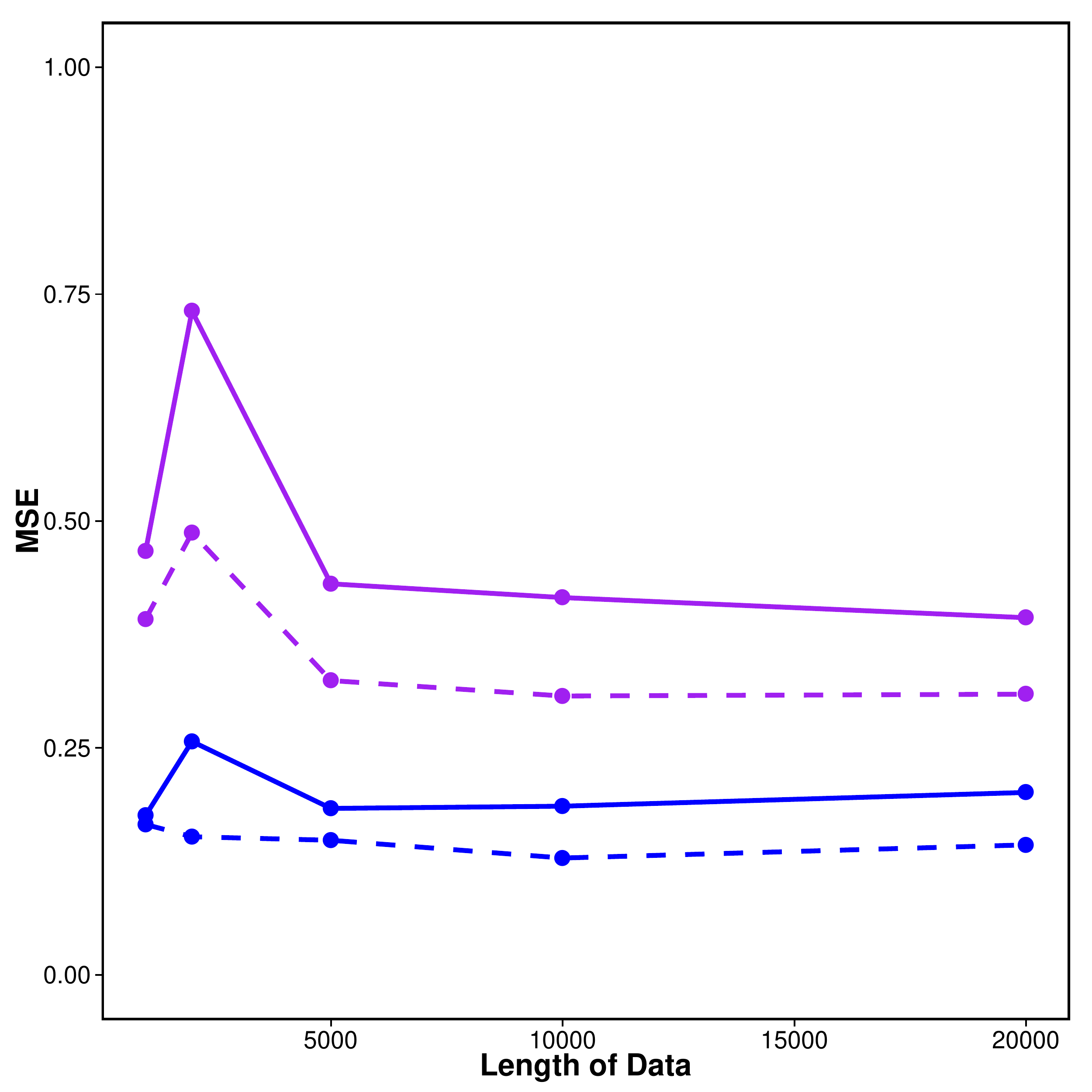}
\end{subfigure}
~
\begin{subfigure}{0.3\linewidth}
\includegraphics[width=\linewidth, height = 5.5cm]{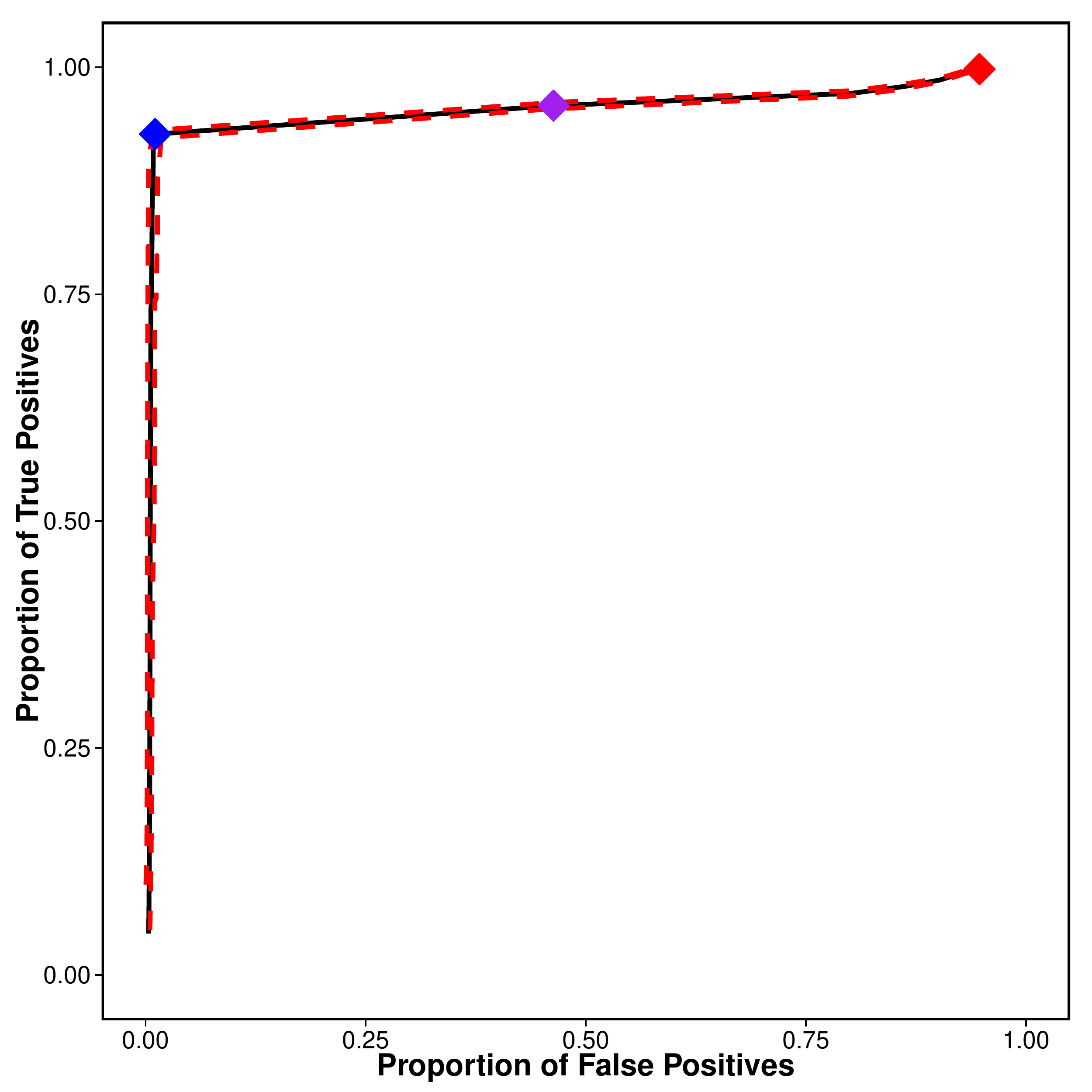}
\end{subfigure}
\caption{Results for the true model. (a) Average minimum (red, dashed) and maximum (grey, dot-dashed) optimal penalty values in comparison to popular penalty terms in the literature. Full lines from top to bottom are the SIC, Hannan-Quinn and AIC penalty values. (b) MSE for the mean (solid) and the standard deviation (dashed) 
when different penalty terms are used; SIC (Blue) and Hannan-Quinn (Purple). (c) Proportion of true positives against the proportion of false positives for $n=10,000$. Red point: AIC, blue point: 
SIC, purple point: Hannan-Quinn.  The red dashed lines show the confidence bounds.  
The top row is the results for fixed changepoints ($m = 10$), the middle row is sublinear changepoints ($m = \sqrt{n}/4$) and the bottom row is linear changepoints ($m = n/100$).}
\label{fig:meanvar}
\end{figure}

The results we obtain when we analysed data from the true model are given in Figure \ref{fig:meanvar}. The left-hand column shows the average range of $\beta$ values that would give estimates of the true number of
changepoints as a function of data size for the three scenarios for the number of changepoints.  
It can be seen that, in this example, when we have 10 changepoints in the data the optimal value of the penalty lies in a wide interval which increases with data size.  
In this case we can see that the AIC, SIC and Hannan-Quinn penalty values will all tend to over-fit the data and hence find too many changes.  
In comparison when the number of changepoints increases with the amount of data, we see that the interval in which the optimal penalty value lies decreases as the length of the data increases. 
In this case the SIC underestimates the number of changes whereas the AIC and Hannan-Quinn penalty term will both tend to overestimate the number of changes. When there is a sublinear number of changepoints the optimal penalty value lies in a smaller interval than it did when there was a fixed number of changes.  In this case the SIC, AIC and Hannan-Quinn penalty all overestimate the number of changepoints.

The MSE for the SIC (blue) and Hannan-Quinn (purple) penalties can be seen in the middle column of Figure \ref{fig:meanvar}.  
The SIC penalty outperforms the Hannan-Quinn penalty in all cases. In all cases the MSE for the AIC penalty term was much larger than the other two penalties and thus not shown in this plot.  

The results for the accuracy of estimating the position of changepoints, for $n = 10,000$, is shown in Figure \ref{fig:meanvar_ROC}; the results are similar for other data lengths.  
It is clear to see that both the AIC and Hannan-Quinn penalty detect a lot of false positive changepoints.  The SIC performs well for all three cases of the numbers of changepoints. 

\begin{figure}
\centering
\begin{subfigure}{0.45\linewidth}
\caption{}\includegraphics[width=\linewidth, height = 5cm]{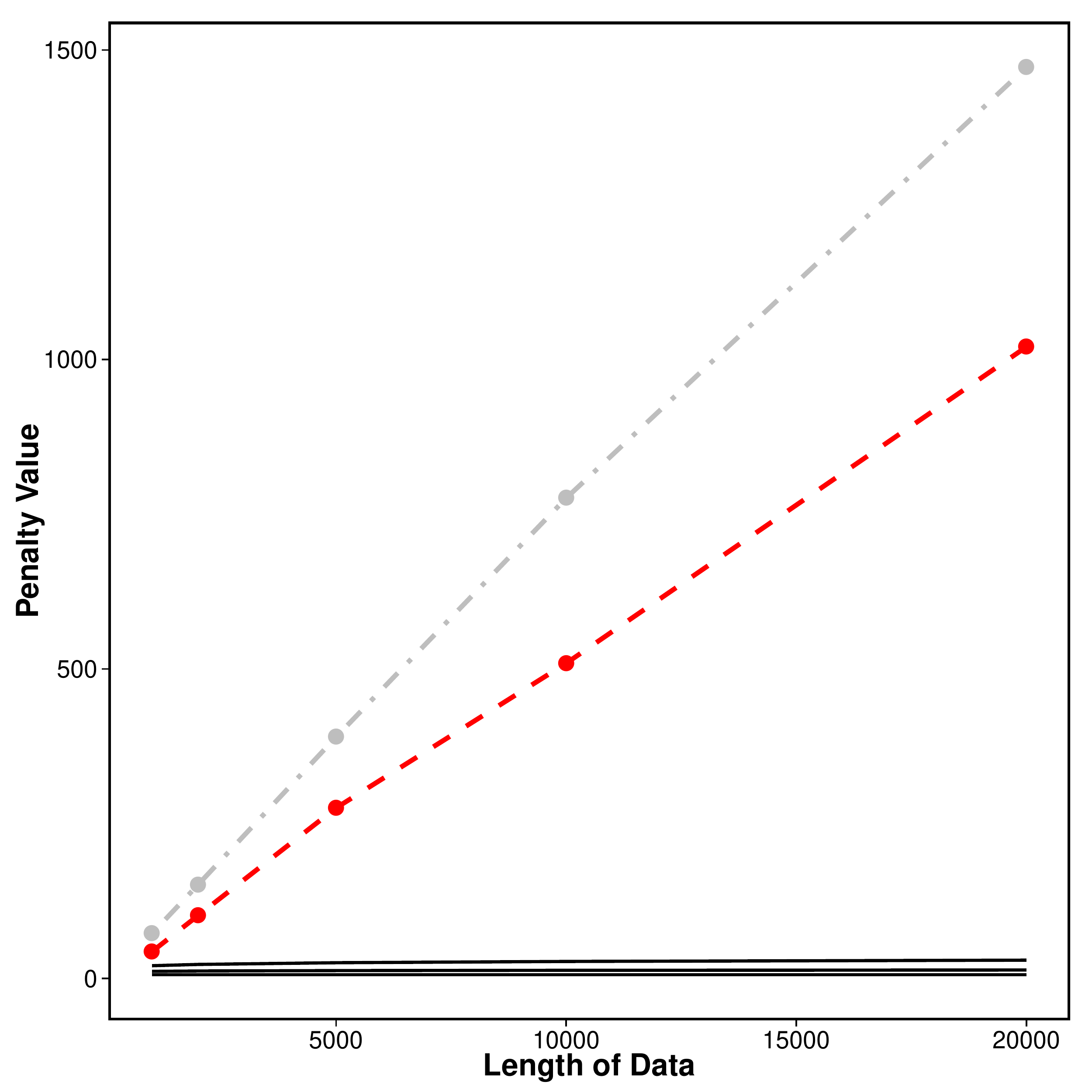} \label{fig:meanvar_mis_res}
\end{subfigure}
~
~
\begin{subfigure}{0.45\linewidth}
\caption{}\includegraphics[width=\linewidth, height = 5cm]{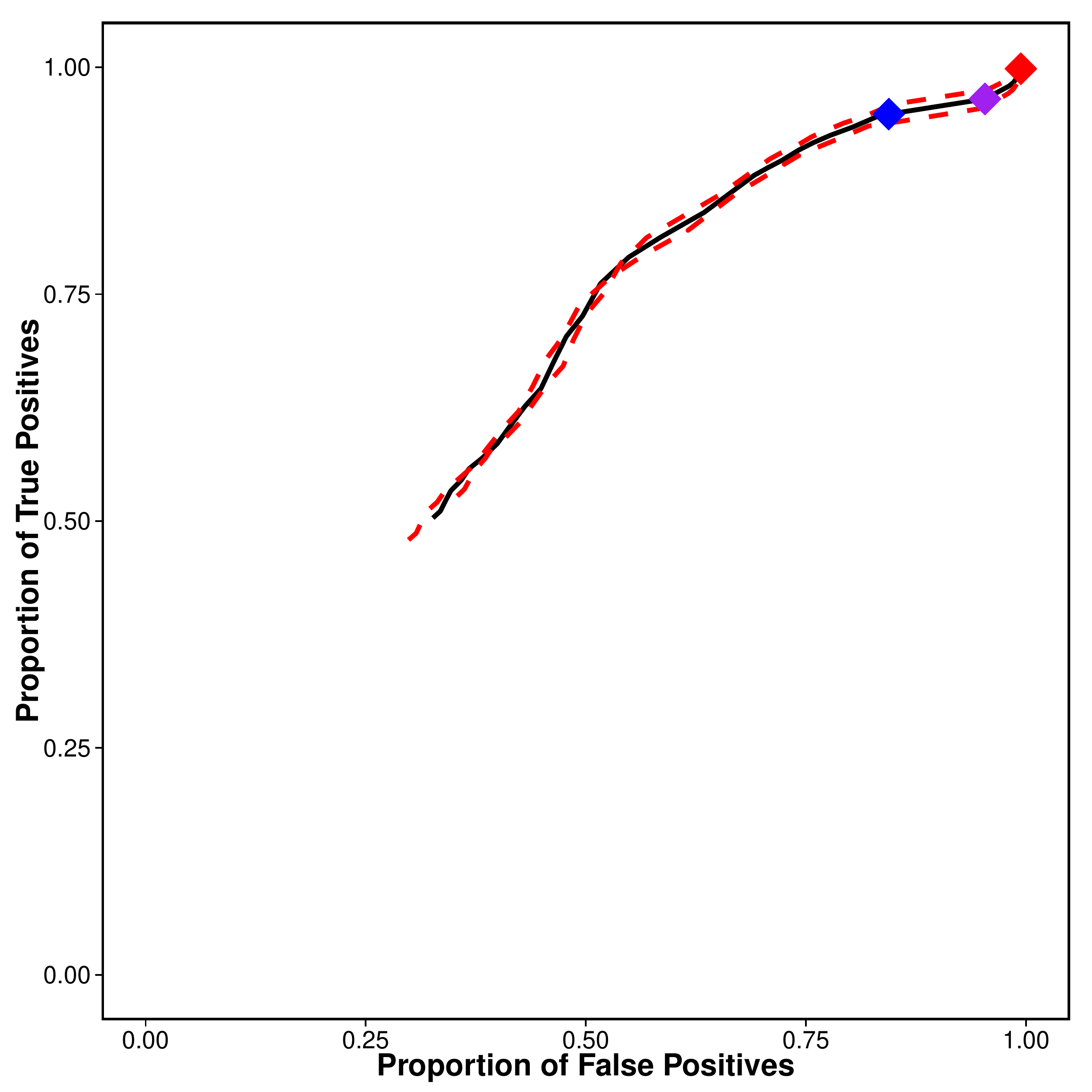}
\label{fig:meanvar_mis_ROC}
\end{subfigure}

\begin{subfigure}{0.45\linewidth}
\includegraphics[width=\linewidth, height = 5cm]{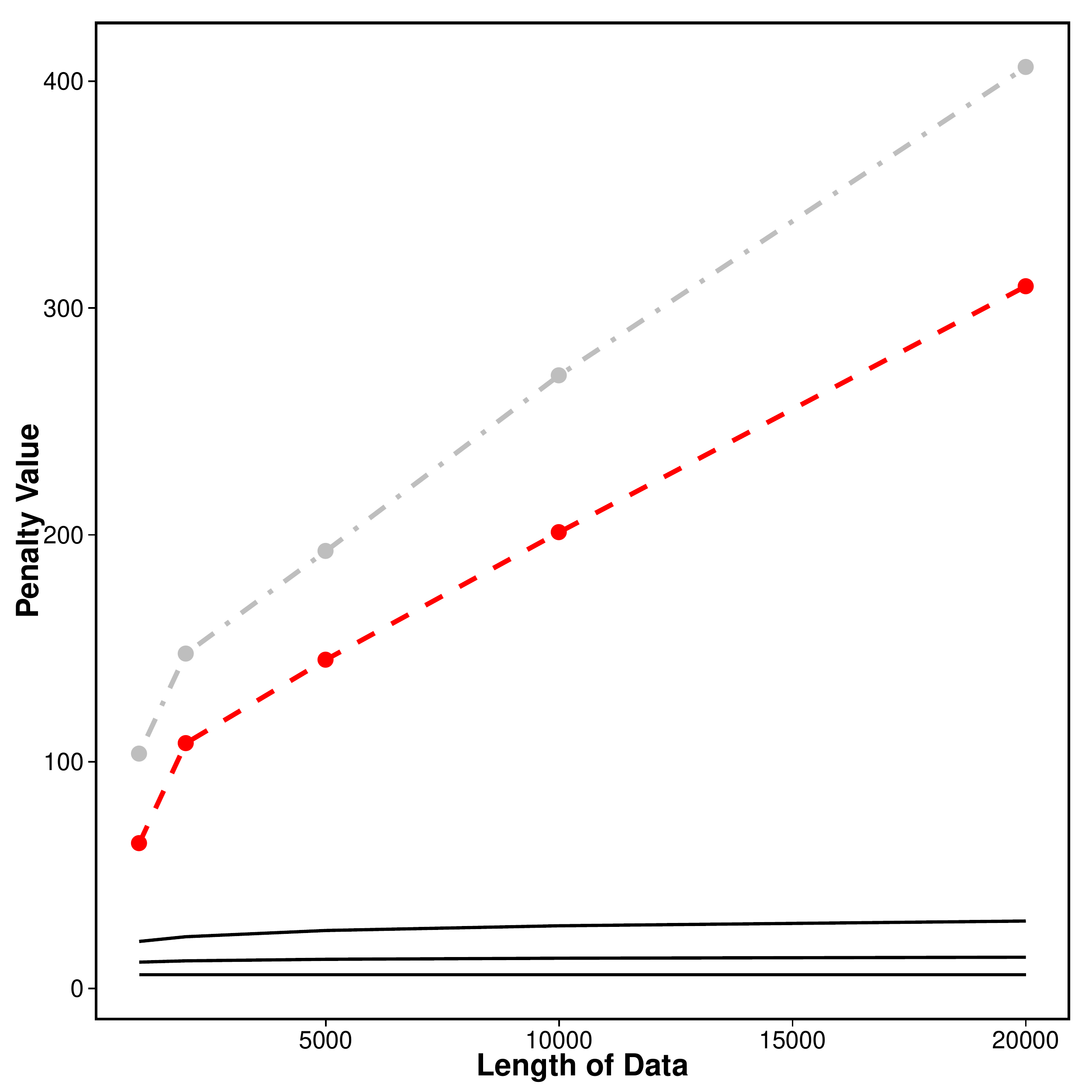}
\end{subfigure}
~
~
\begin{subfigure}{0.45\linewidth}
\includegraphics[width=\linewidth, height = 5.5cm]{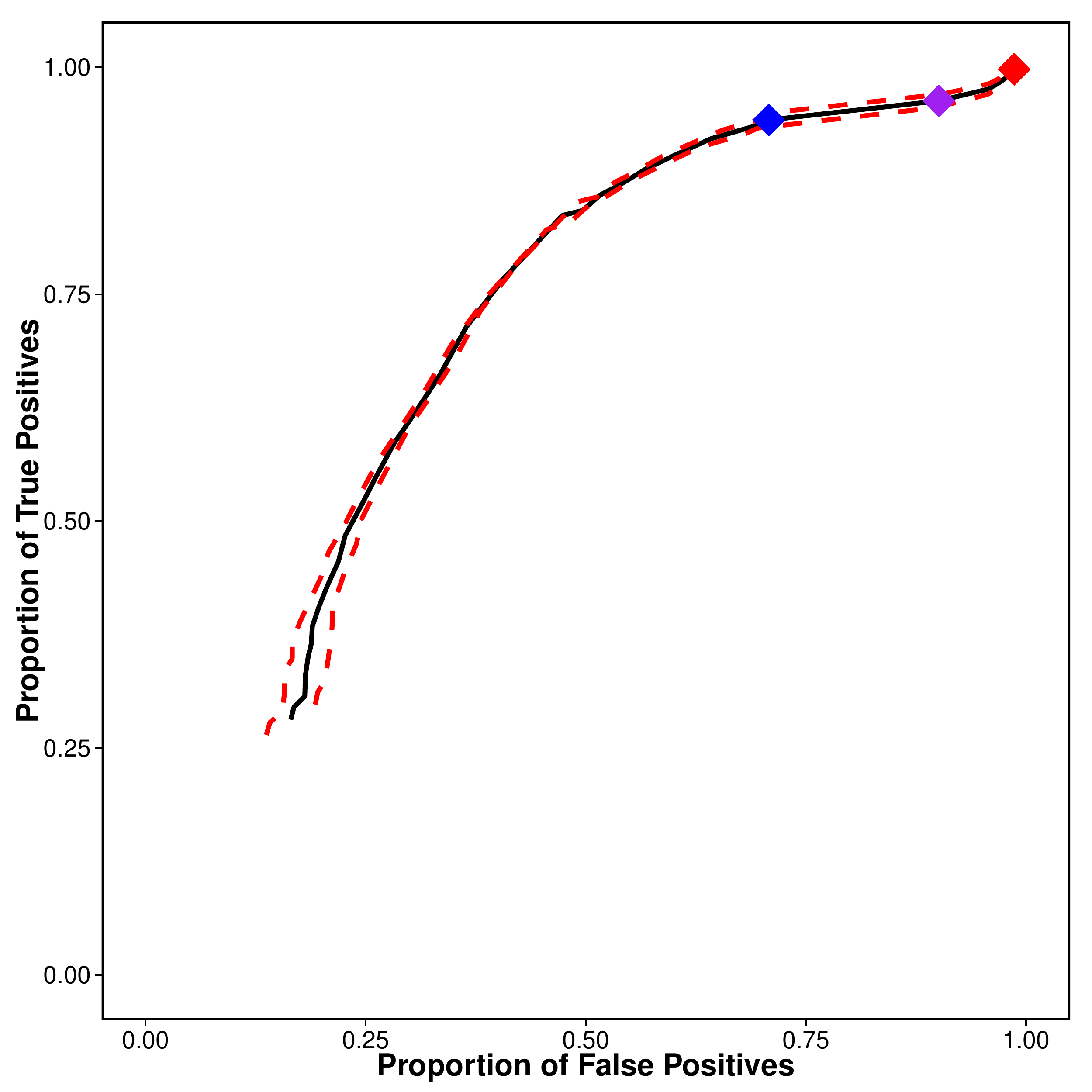}
\end{subfigure}

\begin{subfigure}{0.45\linewidth}
\includegraphics[width=\linewidth, height = 5cm]{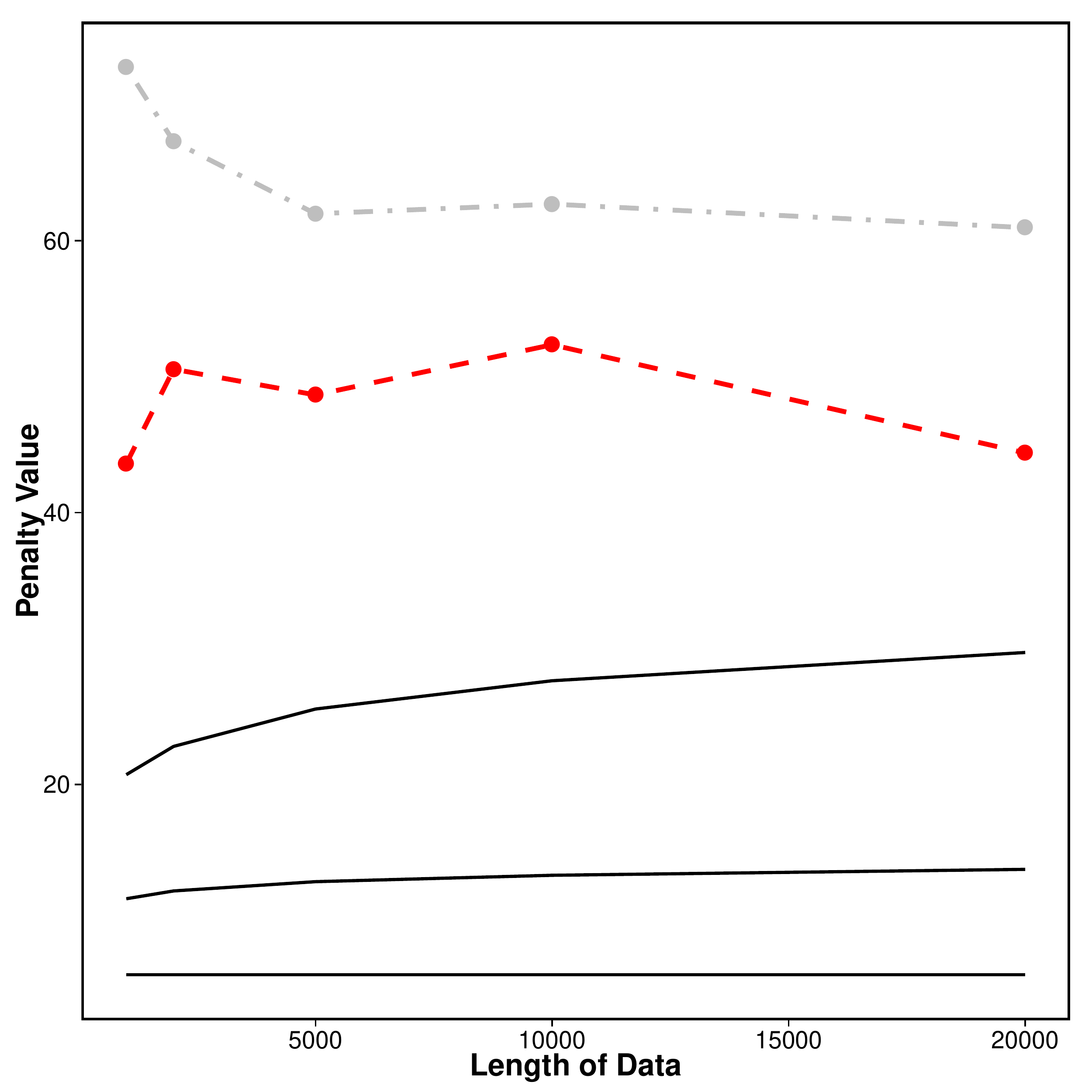}
\end{subfigure}
~
~
\begin{subfigure}{0.45\linewidth}
\includegraphics[width=\linewidth, height = 5.5cm]{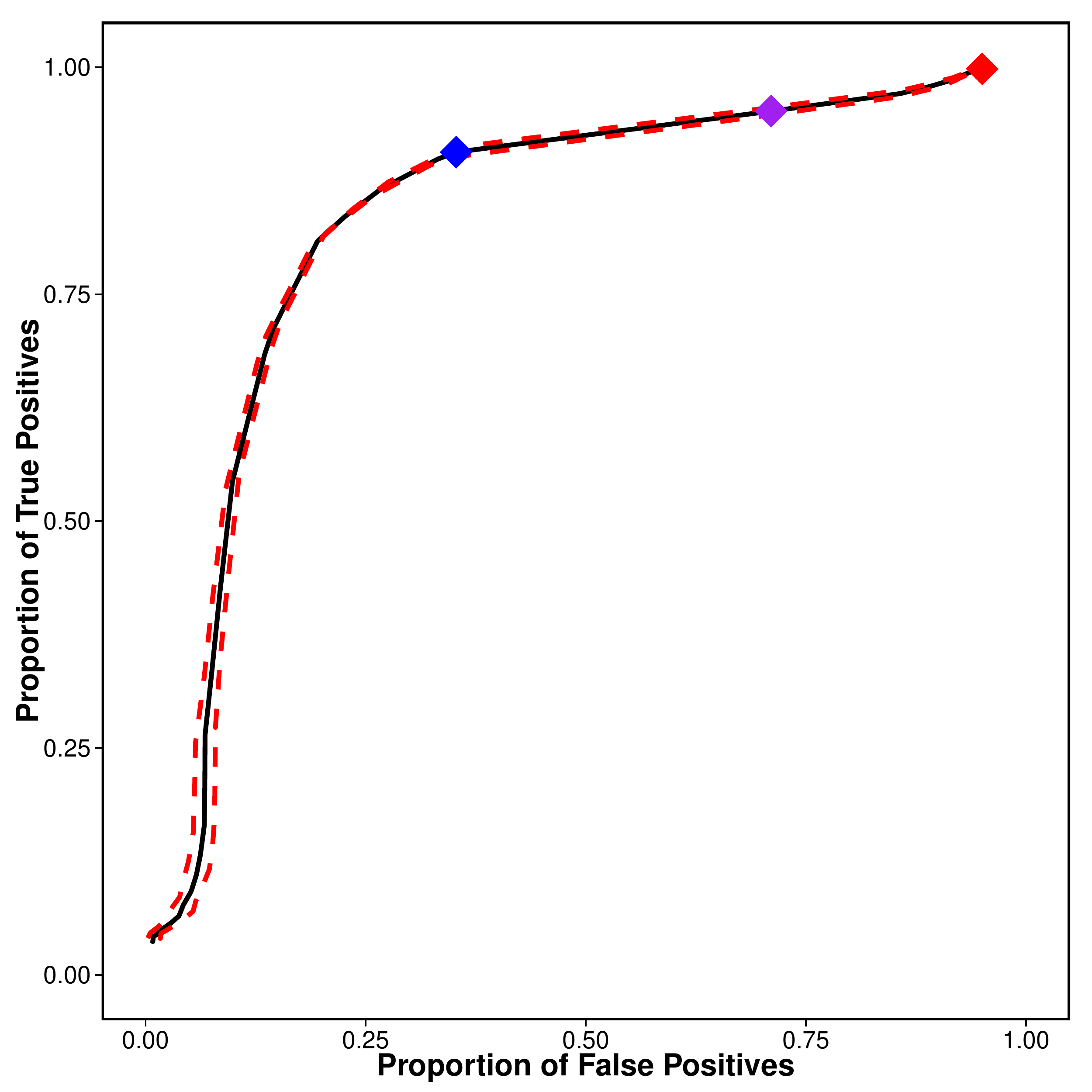}
\end{subfigure}

\caption{Results for the misspecified model scenario.  (a) Average minimum (red, dashed) and maximum (grey, dot-dashed) optimal penalty values in comparison to popular penalty terms in the literature. 
(b) Proportion of true positives against the proportion of false positives for $n=10,000$. Red point: AIC, blue point: 
SIC, purple point: Hannan-Quinn.  The red dashed lines show the confidence bounds.  
The top row is the results for fixed changepoints ($m = 10$), the middle row is sublinear changepoints ($m = \sqrt{n}/4$) and the bottom row is linear changepoints ($m = n/100$).}
\label{fig:meanvar_mis}
\end{figure}

We now look at the case where we have the mis-specified model.  
As above we look at the average of the range of $\beta$ values needed to estimate the correct number of changepoints, and also at the number of true and false positives as we vary $\beta$. We do not consider the mean
square error of the parameter estimates, as there is no direct correspondence between the true and estimated parameters because of the model mis-specification.
  The results can be seen in Figure \ref{fig:meanvar_mis}.  
It is obvious from these results that the optimal penalty value, in terms of correctly estimating the number of changepoints, is much greater than that for the correctly specified model.  
It is also much larger than any of  SIC, AIC and Hannan-Quinn.  
From the accuracy plots we can see that 
none of the penalty terms perform well, with them all detecting a large number of false positives.  

\section{Application to well-log data} \label{real_data}

We now demonstrate CROPS for detecting changes in well-log data 
\cite[]{ORuanaidh96}.  This data set contains information about different rock strata obtained by recording measurements of nuclear magnetic response as a probe is lowered
down a bore-hole into the earth's surface.
The ability to predict changes in rock type is useful for drilling as it allows for the drilling pressure to be adjusted to avoid blow-outs.  The original data set contains outliers which we have removed before analysing. 

We primarily wish to detect a change in the mean of the process. However to allow for not knowing an appropriate common variance of the noise, we 
apply our changepoint detection method with a range of penalty values using the change in mean and variance cost function. Note that a simple change in mean, or change in mean and variance,
is an over-simplistic model for this data, as the mean of the process appears to vary slowly between the abrupt changes.

\begin{figure}[t]
\centering
\includegraphics[width=\linewidth]{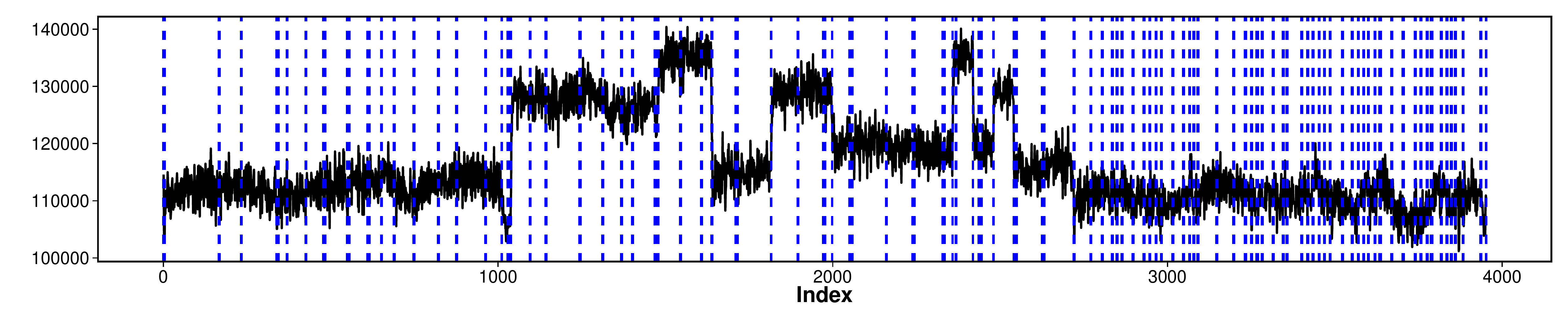}
\caption{Segmentation of the well-oil data using the SIC penalty term.}
\label{Fig:oil_SIC}
\end{figure}

The segmentation using the SIC penalty term is shown in Figure \ref{Fig:oil_SIC}.  
It can be seen that in this example using this penalty term massively over fits the data; 218 segments are detected. This suggests that these
penalty values are much too small for this application, due to the over-simplistic model being fit to data within a segment. 

An alternative way to choosing the number of changepoints, or equivalently the penalty value, has been suggested by \cite{Lavielle2005}.
This involves plotting the un-penalised cost against the number of segments, $m$.  
Initially as we increase $m$ we are likely to be detecting true changepoints, and these should
lead to a substantial decrease in the cost.  As we detect more changes these will eventually become false positives, and 
we would expect that detecting a false positive will not lower the cost as much.  
Thus \cite{Lavielle2005} suggests choosing the point where the decrease in cost due to detecting further changepoint noticeably changes. This can be thought of as looking for an ``elbow'' in the plot of the unpenalised
cost versus $m$. In practice such an approach may suggest a plausible range of values for $m$, and these could then be considered in turn as alternative segmentations. 

\begin{figure}[t]
\centering

\begin{subfigure}{0.47\linewidth}
\includegraphics[width=\linewidth]{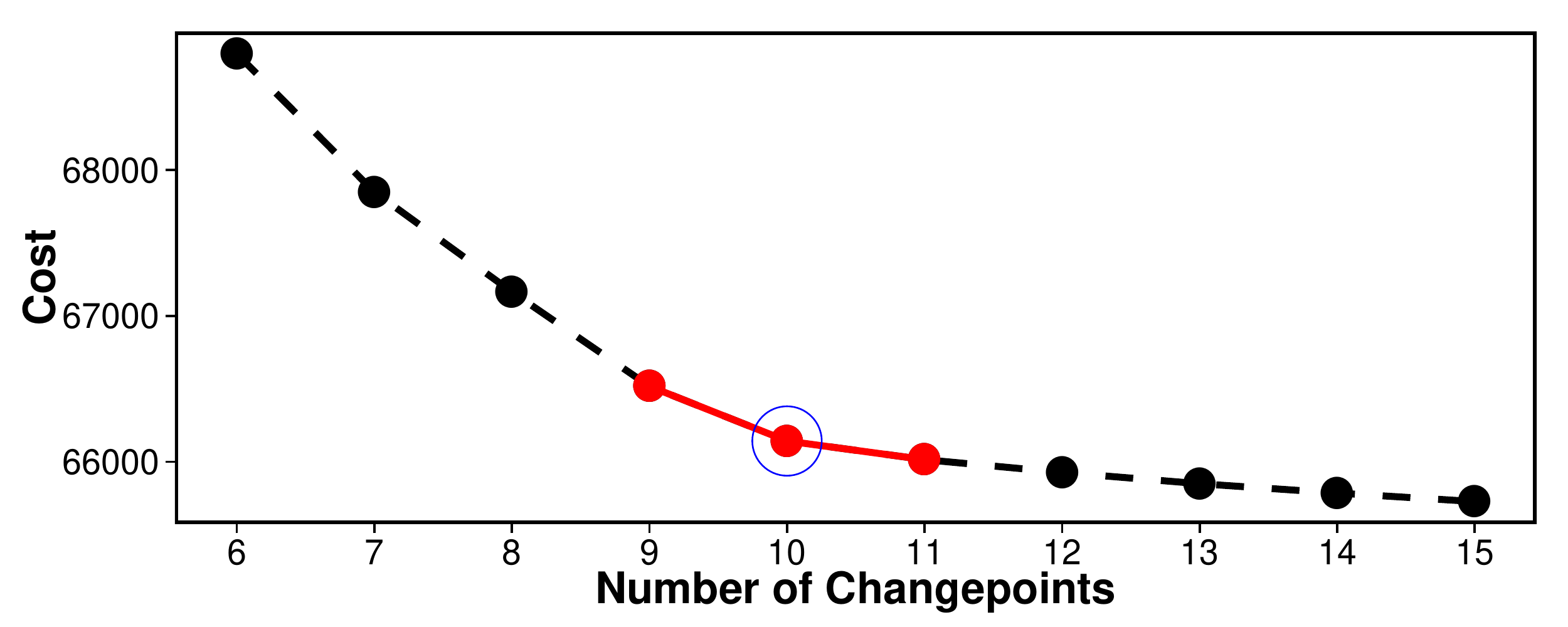}
\end{subfigure}
~
\begin{subfigure}{0.47\linewidth}
\includegraphics[width=\linewidth]{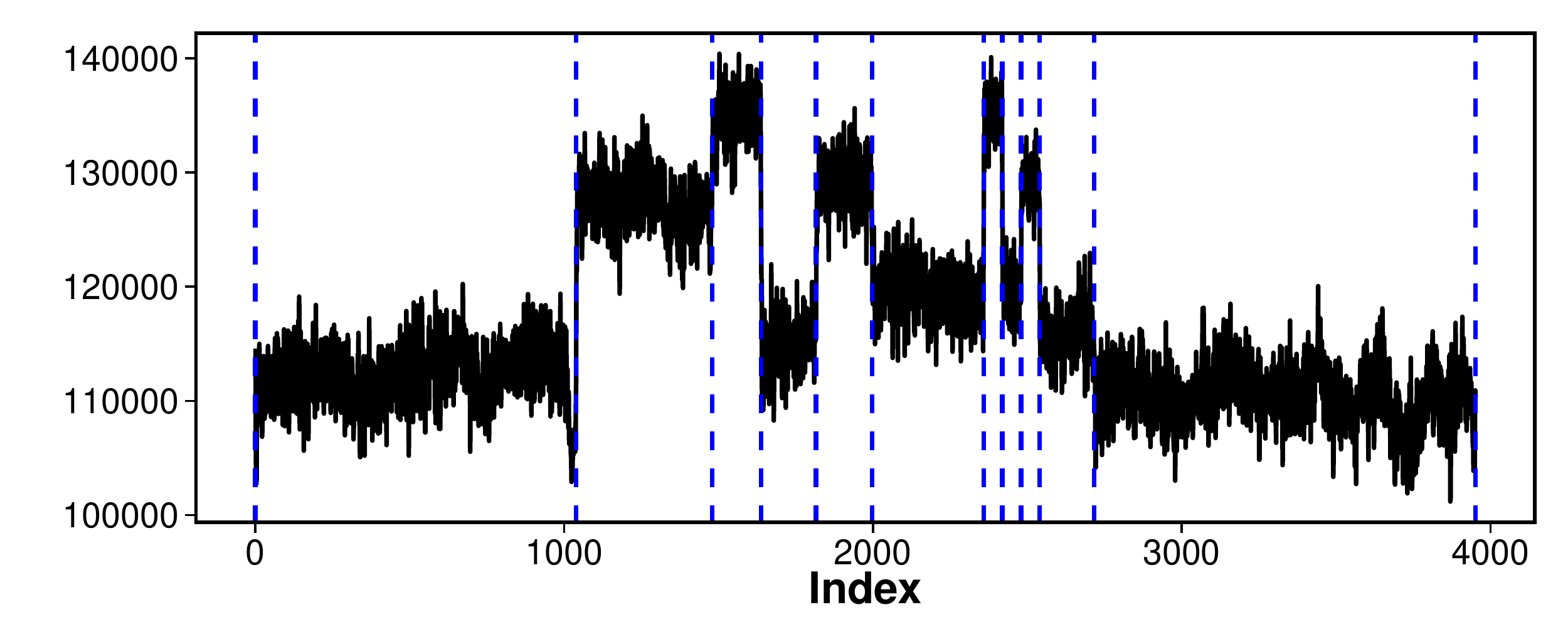}
\end{subfigure}

\begin{subfigure}{0.47\linewidth}
\includegraphics[width=\linewidth]{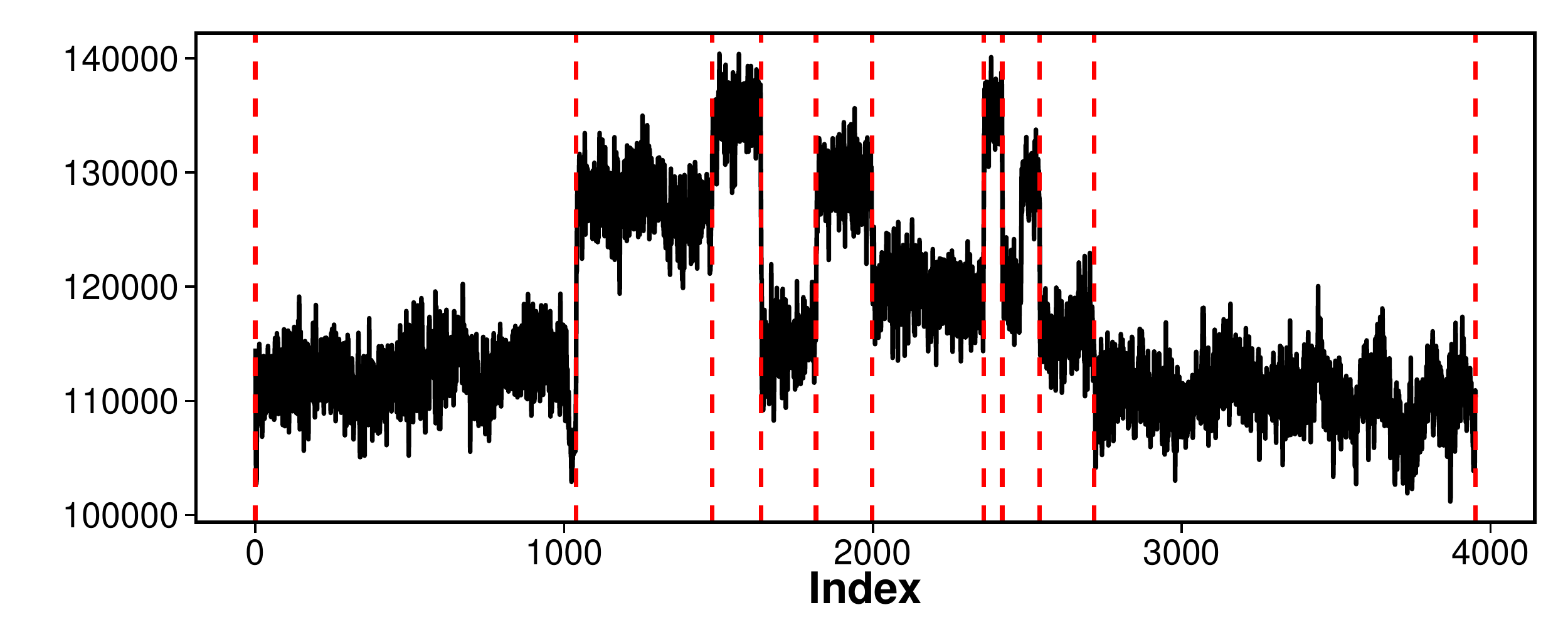}
\end{subfigure}
~
\begin{subfigure}{0.47\linewidth}
\includegraphics[width=\linewidth]{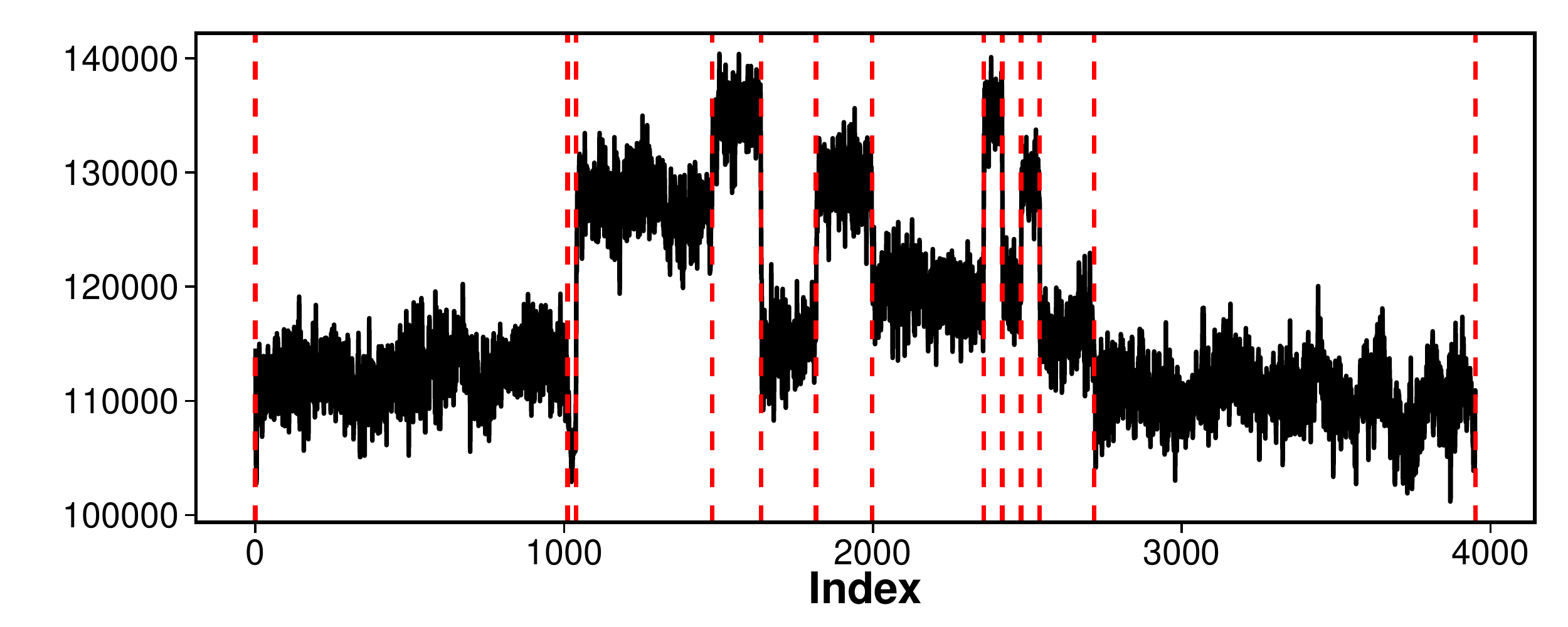}
\end{subfigure}

\caption{Well-oil data. Top left: Resulting penalised likelihood against the number of changepoints, top right: Plot of the well-log data with a segmentation with 11 segments, bottom left: segmentation with 10 segments and bottom right: segmentation with 12 segments.}
\label{Fig:oil_ans}
\end{figure}

The plot of the unpenalised cost against the number of segments for this example is shown in Figure \ref{Fig:oil_ans}.  
The blue circle indicates a point, which by eye, could be described as being the ``elbow'' and the red points indicate the points near the elbow, which we could also have chosen.  
The resulting segmentations can also be seen in Figure \ref{Fig:oil_ans}. 
Of these, the segmentation of the data into 12 segments looks most sensible.

%% file: discussion.tex
%
%
%
%

\section{Discussion}\label{sec:discussion} 
In this paper we have developed a method, CROPS, to obtain the optimal segmentations of data, based on minimising a penalised cost function, for a range of penalty values.  
For many applications, we believe this is a more appropriate approach to segmenting data than just using a single choice of penalty, such as SIC. In particular, whilst
default choices can work well if we have an accurate model for the data within each segment, we have shown that they lack robustness, and can produce poor segmentations, 
in the presence of model mis-specification. We have observed such issues in both a simulation study, and when analysing the well-log data.

Minimising the penalised cost function for a range of penalty values is one way of producing a number of different ways of segmenting data, each with a different number of segments.
As such, this approach is an alternative to the Segment Neighbourhood search method, which outputs the optimal segmentation as the number of segments is varied across a suitably chosen range.
The advantage of the new approach is one of computational speed, which benefits from the fact that minimising the penalised cost function is a simpler problem to solve than minising the
cost function under a constraint on the number of changepoints, the problem that Segment Neighbourhood solves. In our simulations, CROPS was up to two orders of magnitude quicker
than Segment Neighbourhood. One advantage of Segment Neighbourhood is that it produces an optimal segmentation for all numbers of segments in the chosen range, whereas some of these may not be
optimal under the penalised cost function for any penalty value, and hence not found via our new method. However the segmentations we do not recover correspond to, for example, ones where adding an extra
changepoint leads to a larger change in cost than removing a changepoint. It is hard to construct a sensible criteria under which such segmentations would be optimal.

Whilst we have implemented CROPS using PELT to minimise the penalised cost for a given penalty value, any algorithm that can solve the minimisation problem can be used. For some
applications, such as detecting a change in the mean of a uni-variate time-series, we believe that using the OPFP algorithm \cite[]{Maidstone:2014} will lead to substantial further
reductions in computational cost. 

